\documentclass[twocolumn]{aastex63}

\usepackage{graphicx, amsmath}
\usepackage[caption=false]{subfig}
\usepackage{hyperref}
\usepackage{cleveref}
\usepackage{xspace}
\usepackage{rotating}
\usepackage{relsize}
\usepackage{comment}

\newcommand{\lya}{Ly$\alpha$\xspace}

\newcommand{\angstrom}{\mbox{\normalfont\AA\xspace}}

\newcommand{\MgII}{\ion{Mg}{2}\xspace}

\newcommand{\CIV}{\ion{C}{4}\xspace}
\newcommand{\NV}{\ion{N}{5}\xspace}

\newcommand{\civ}{C\,{\sc iv}}
\newcommand{\heii}{He\,{\sc ii}}
\newcommand{\ciii}{C\,{\sc iii}]}

\newcommand{\oii}{[O\,{\sc ii}]}  
\newcommand{\ovi}{O\,{\sc vi}}

\newcommand{\kms}{km s$^{-1}$}

\usepackage{enumitem,amssymb}
\usepackage{float}
\usepackage[utf8]{inputenc}
\usepackage{color}
\usepackage{graphicx} 
\usepackage{amsmath} 
\usepackage{amssymb} 
\usepackage{wasysym}
\usepackage{mathtools}
\usepackage{newtxtext,newtxmath}

\newcommand{\cgsa}{erg s$^{-1}$ cm$^{-2}$ $\angstrom^{-1}$}
\newcommand{\cgs}{erg s$^{-1}$ cm$^{-2}$}

\makeatletter
\DeclareTextCompositeCommand{\r}{OT1}{A}{%
  \leavevmode\vbox{%
    \offinterlineskip
    \ialign{\hfil##\hfil\cr\char23\cr\noalign{\kern-1.15ex}A\cr}%
  }%
}
\makeatother

\graphicspath{{./}{figures/}}
\newcommand{\rv}[1]{#1}
\newcommand{\rvtwo}[1]{#1}

\shorttitle{HPSC-1 LAEs}
\shortauthors{Davis et al}

\begin{document}

\title{HETDEX Public Source Catalog 1 - Stacking 50K Lyman Alpha Emitters \footnote{Based on observations obtained with the Hobby-Eberly Telescope, which is a joint project of the University of Texas at Austin, the Pennsylvania State University, Ludwig-Maximilians-Universit\"at M\"unchen, and Georg-August-Universit\"at G\"ottingen.  The HET is named in honor of its principal benefactors, William P.~Hobby and Robert E.~Eberly.}} 

\author[0000-0002-8925-9769]{Dustin Davis}
\affiliation{Department of Astronomy, The University of Texas at Austin, Austin, TX 78712, USA}

%
%
\author[0000-0002-8433-8185]{Karl Gebhardt}
\affiliation{Department of Astronomy, The University of Texas at Austin, Austin, TX 78712, USA}

\author[0000-0002-2307-0146]{Erin Mentuch Cooper}
\affiliation{Department of Astronomy, The University of Texas at Austin, Austin, TX 78712, USA}

\author[0000-0003-4381-5245]{William P. Bowman}
\affiliation{Department of Astronomy, Yale University, New Haven, CT 06520}

\author{Barbara Garcia Castanheira}
\affiliation{Baylor University, Department of Physics, Waco TX 76798, USA}

\author[0000-0002-0302-2577]{John Chisholm}
\affiliation{Department of Astronomy, The University of Texas at Austin, Austin, TX 78712, USA}

\author[0000-0002-1328-0211]{Robin Ciardullo} \affiliation{Department of Astronomy \& Astrophysics, The Pennsylvania
State University, University Park, PA 16802, USA}
\affiliation{Institute for Gravitation and the Cosmos, The Pennsylvania State University, University Park, PA 16802, USA}

\author[0000-0002-7025-6058]{Maximilian Fabricius}
\affiliation{Max-Planck Institut f\"ur extraterrestrische Physik, Giessenbachstrasse 1, 85748 Garching, Germany}
\affiliation{Universit{\"a}ts-Sternwarte, Fakult{\"a}t f{\"u}r Physik, Ludwig-Maximilians Universit{\"a}t M{\"u}nchen, Scheinerstr. 1, 81679\ M{\"u}nchen, Germany}

\author[0000-0003-2575-0652]{Daniel J. Farrow}
\affiliation{Centre of Excellence for Data Science, Artificial Intelligence and Modelling (DAIM), University of Hull, Cottingham Road, Kingston-upon-Hull HU6 7RX, UK}
\affiliation{E. A. Milne Centre for Astrophysics, University of Hull, Cottingham Road, Kingston-upon-Hull HU6 7RX, UK}

\author[0000-0001-8519-1130]{Steven L. Finkelstein}
\affiliation{Department of Astronomy, The University of Texas at Austin, Austin, TX 78712, USA}

\author[0000-0001-6842-2371]{Caryl Gronwall}
\affiliation{Department of Astronomy \& Astrophysics, The Pennsylvania
State University, University Park, PA 16802, USA}
\affiliation{Institute for Gravitation and the Cosmos, The Pennsylvania State University, University Park, PA 16802, USA}

\author[0000-0003-1530-8713]{Eric Gawiser}
\affiliation{Department of Physics \& Astronomy, Rutgers, The State University of New Jersey, Piscataway, NJ 08854, USA}

\author[0000-0001-6717-7685]{Gary J. Hill}
 \affiliation{McDonald Observatory, The University of Texas at Austin, Austin, TX 78712, USA}
\affiliation{Department of Astronomy, The University of Texas at Austin, Austin, TX 78712, USA}

\author[0000-0003-1008-225X]{Ulrich Hopp}
\affiliation{Universit{\"a}ts-Sternwarte, Fakult{\"a}t f{\"u}r Physik, Ludwig-Maximilians Universit{\"a}t M{\"u}nchen, Scheinerstr. 1, 81679\ M{\"u}nchen, Germany}
\affiliation{Max-Planck Institut f\"ur extraterrestrische Physik, Giessenbachstrasse 1, 85748 Garching, Germany}

\author[0000-0002-1496-6514]{Lindsay R. House}
\affiliation{Department of Astronomy, The University of Texas at Austin, Austin, TX 78712, USA}

\author[0000-0002-8434-979X]{Donghui Jeong}
 \affiliation{Department of Astronomy \& Astrophysics, The Pennsylvania State University, University Park, PA 16802, USA}
 \affiliation{Institute for Gravitation and the Cosmos, The Pennsylvania State University, University Park, PA 16802, USA}


\author[0000-0002-0417-1494]{Wolfram Kollatschny}
\affiliation{Institut f\"{u}r Astrophysik, Universit\"{a}t G\"{o}ttingen, Friedrich-Hund-Platz 1, 37077 G\"{o}ttingen, Germany}

 \author[0000-0002-0136-2404]{Eiichiro Komatsu}
 \affiliation{Max-Planck-Institut f\"ur Astrophysik, Karl-Schwarzschild-Str. 1, 85748 Garching, Germany}
 \affiliation{Kavli Institute for the Physics and Mathematics of the Universe (WPI),
Todai Institutes for Advanced Study, the University of Tokyo, Kashiwanoha, Kashiwa, Chiba 277-8583, Japan}

\author[0000-0001-5561-2010]{Chenxu Liu}
\affiliation{South-Western Institute for Astronomy Research, Yunnan University, Kunming, Yunnan, 650500, People’s Republic of China}
\affiliation{Department of Astronomy, The University of Texas at Austin, Austin, TX 78712, USA}

\author[0000-0002-6907-8370]{Maja Lujan Niemeyer}
\affiliation{Max-Planck-Institut f\"ur Astrophysik, Karl-Schwarzschild-Str. 1, 85748 Garching, Germany}

\author[0000-0001-8419-3062]{Alberto Saldana-Lopez}
\affiliation{Department of Astronomy, University of Geneva, 51 Chemin Pegasi, 1290 Versoix, Switzerland}

\author[0000-0002-6186-5476]{Shun Saito}
\affiliation{Institute for Multi-messenger Astrophysics and Cosmology, Department of Physics,
Missouri University of Science and Technology, 
1315 N. Pine St., Rolla MO 65409, USA}
\affiliation{Kavli Institute for the Physics and Mathematics of the Universe (WPI),
Todai Institutes for Advanced Study, the University of Tokyo, Kashiwanoha, Kashiwa, Chiba 277-8583, Japan}

\author[0000-0001-7240-7449]{Donald P. Schneider}
\affiliation{Department of Astronomy \& Astrophysics, The Pennsylvania State University, University Park, PA 16802, USA}
\affiliation{Institute for Gravitation and the Cosmos, The Pennsylvania State University, University Park, PA 16802, USA}

\author[0000-0003-4044-5357]{Jan Snigula}
\affiliation{Universit{\"a}ts-Sternwarte, Fakult{\"a}t f{\"u}r Physik, Ludwig-Maximilians Universit{\"a}t M{\"u}nchen, Scheinerstr. 1, 81679\ M{\"u}nchen, Germany}
\affiliation{Max-Planck Institut f\"ur extraterrestrische Physik, Giessenbachstrasse 1, 85748 Garching, Germany}

\author[0000-0002-7327-565X]{Sarah Tuttle}
\affiliation{Department of Astronomy, University of Washington, Seattle, 3910 15th Ave NE, Room C319, Seattle WA 98195-0002}

\author[0000-0002-4974-1243]{Laurel H. Weiss}
\affiliation{Department of Astronomy, The University of Texas at Austin, Austin, TX 78712, USA}

\author{Lutz Wisotzki}
\affiliation{Leibniz-Institut f\"ur Astrophysik Potsdam (AIP), An der Sternwarte 16, 14482 Potsdam, Germany}

\author[0000-0003-2307-0629]{Gregory Zeimann}
\affiliation{Hobby Eberly Telescope, University of Texas, Austin, Austin, TX, 78712, USA}

\defcitealias{Gebhardt+2021}{KG21}
\defcitealias{Davis2021}{DD21}

\begin{abstract}

We describe the ensemble properties of the $1.9 < z < 3.5$ Lyman Alpha Emitters (LAEs) found in the HETDEX survey's first public data release, HETDEX Public Source Catalog 1 \citep{Cooper_2022}.  
Stacking the low-resolution ($R \sim$ 800) spectra greatly increases the signal-to-noise ratio, revealing spectral features otherwise hidden by noise, and we show that the stacked spectrum is representative of an average member of the set. The flux limited, \lya\ signal-to-noise ratio restricted stack of 50K HETDEX LAEs shows the ensemble biweight ``average" $z \sim 2.6$ LAE to be a blue (UV continuum slope $\sim -2.4$ and E(B-V) $< 0.1$), moderately bright (M$_{\text{UV}} \sim -19.7$) star forming galaxy with strong \lya\ emission (log $L_{Ly\alpha}$ $\sim$ 42.8 and $W_{\lambda}$(\lya) $\sim$ 114\AA), and potentially significant leakage of ionizing radiation. The restframe UV light is dominated by a young, metal poor stellar population with an average age 5-15 Myr and metallicity of 0.2-0.3 Z$_{\odot}$. 
\end{abstract}


\section{Introduction}\label{sec:intro} 

The Hobby-Eberly Telescope Dark Energy Experiment \citep[HETDEX,][]{Gebhardt+2021,Hill+21} is a multi-year, untargeted, low-resolution ($R \sim$ 800) spectroscopic survey conducted with the Hobby-Eberly Telescope \citep[HET,][]{Ramsey1998, Hill+21} and the Visible Integral-field Replicable Unit Spectrograph \citep[VIRUS;][]{Hill+21}. Object spectra are identified post-observation via examination of the spatial and spectral clustering of the individual optical spectra from the $\sim$35K fibers in the array (\S \ref{sec:hetdex}). The HETDEX Public Source Catalog 1, hereafter, HPSC-1 \citep{Cooper_2022} contains spectra for more than 200K objects from the first three years of HETDEX observations. Within this data set are more than 50K Lyman Alpha Emitting (LAE) galaxies at $1.9 < z < 3.5$ identified by their \lya\ emission lines, which HETDEX is specifically designed to detect. Using the 3D positions of these LAEs, which serve as biased mass tracers, HETDEX aims to constrain the Hubble Parameter, $H(z)$, and the Angular Diameter Distance, $D_{A}(z)$, at $z\sim$2.4 to better than 1\% accuracy \citep{Gebhardt+2021}. The success of this primary science goal is predicated on the accurate measurement of the redshifts, and hence the correct identification of the \lya\ emission line, of the roughly one million LAEs out of the several million total galaxies and other astrophysical sources expected to be observed over the course of the HETDEX survey. The spectra obtained from the untargeted observations are exploited to this end \citep{Davis_2022}.

The HETDEX spectral range is 3500-5500~\angstrom\ with $\lambda /\Delta \lambda  \sim 800$ \citep{Hill+21}. Most, $\sim$ 80\%, of all HETDEX emission line detections (or $\sim$ 70\% of all HETDEX objects when including those identified from their continuum rather than emission lines) are in spectra without significantly detected continua and thus provide very little extra information beyond their emission line specified redshifts. However, with the classifications and redshifts of these objects known to high precision and with little misidentification, especially for \lya \citep[only $\sim$2\% of all identified \lya\ emission lines are not \lya][]{Davis_2022}, it becomes practical to stack the spectra to explore the average properties of sub-populations of galaxies (specifically LAEs, for this work). 
Throughout this paper, we use the biweight measure of the central location \rv{\footnote{\rv{a.k.a the ``biweight location". Also referred to as the ``biweight average" or just the ``biweight" in this and other works.}}} \citep{Beers_1990} as a measure of the average properties in the stack. This is very similar to the median in the limit of a large sample size, such as presented here (see also \S \ref{stacking}), \rv{but provides an improved robustness to outliers, particularly when the sample distribution is non-Gaussian. When the sample is Gaussian-like, the difference is $\ll$ 1\%. For simple comparisons and convenience, the mean or median and standard deviation may also be used, but are explicitly identified.}

Here we define LAE to mean $1.88 < z < 3.52$ galaxies that do not host Active Galactic Nuclei (AGN\null). With the HETDEX emission line search, these galaxies are nearly all ($\gtrsim$ 96\%) classical Lyman Alpha Emitters by definition of a 20~\angstrom\ or greater rest-frame equivalent width \lya\ emission \citep{Gronwall_2007,Adams_2011}. Only a relative few these ``LAEs" have sufficiently bright continua to be selected as a Lyman Break Galaxies \citep[LBGs,][]{Guhathakurta_1990,Madau_1996,Steidel_1996,Shapley_2003}. Generally, the LAEs in the HETDEX redshift window are compact, low-metallicity, rapidly star forming galaxies \citep[][and many others]{Gawiser_2007,Nilsson_2009,Finkelstein_2010}. 

The stacking of spectra, whether for LAEs or other phenomena, is not new \citep[][and many others]{Green_1996, hu_cowie_capak_kakazu_2005, Balestra_2010, Berry_2012, Jaskot_2014, Grazian_2016, Steidel_2018,Feltre_2020b}. HETDEX is unusual in its lack of preselection coupled with the large number of available spectra, which will number in the millions by the end of the survey. High-quality spectral observations of individual $z>2$ galaxies remain extremely rare and limit our ability to extrapolate to a population description, where stacking, by its statistical nature, describes population features. 

The large number of spectra from the observations included in HPSC-1 \citep[which contains roughly 1/4 of the planned 540~deg$^{2}$ sky coverage of the full HETDEX survey;][]{Gebhardt+2021,Cooper_2022} provide significant statistical leverage. Given the untargeted nature of the HETDEX observations and assuming the spectra are down-selected without significant restriction on their on-sky locations, the stacks are largely unbiased with respect to the LAEs' environments. Additionally, these statistically large stacks marginalize over galaxy orientation, dust geometry, star formation stochasticity, and lines of sight through the IGM\null. That last marginalization allows for a more straightforward application of IGM attenuation corrections, which are inherently statistical averages \citep{Meiksin,Byrohl_2020,bassett_2022}, while the other properties embody the random variability in the galaxy population and, in particular, their radiative transfer processes \citep{Heckman_2011,Leitet_2013,Steidel_2018, saldana-lopez_2022,Smith_2022}. 
Further, with 10K-50K spectra in each of the stacks in this work, the signal-to-noise ratio (SNR) is boosted by a factor of 100 or more. This makes possible the measure of signal that is otherwise buried in noise for individual spectra.

In this work, we stack subsamples of the HPSC-1 spectra to explore the basic ensemble properties of $1.9 < z < 3.5$ LAEs and preview future analyses using even larger collections of HETDEX spectra.  

The remainder of this paper is organized as follows:  Section \ref{sec:hetdex} briefly describes the HETDEX observations and reduction pipeline. Section \ref{sec:methods} provides an overview of the selection and stacking mechanics. Section \ref{sec:analysis} presents the results and discusses the various explored properties of the data sets.

Throughout the paper, the Planck 2018 cosmology \citep{Planck2018} with $\Omega_{\text{m}}$= 0.31 and $H_0$ = 67.7 $\mathrm{km~s^{-1}~Mpc^{-1}}$ is assumed. All magnitudes are in the AB system.\\


\section{Observations and Data Selection} \label{sec:hetdex}

For a more thorough and complete description of the HETDEX observations, data reduction, and the particular composition of HPSC-1, we refer the readers to \cite{Gebhardt+2021} and \cite{Cooper_2022}. 

In brief, HETDEX is a multi-year spectroscopic survey conducted at the McDonald Observatory with the 10\,m Hobby-Eberly Telescope. Observations began in 2017 and will continue through 2024 with a planned coverage of some 540 deg$^{2}$ in two large fields pointing out of the Galatic plane, one centered near 193$^{\circ}$+53$^{\circ}$ and the other near 22$^{\circ}$+0$^{\circ}$. Pointings within the two fields are untargeted, and the spectra are collected with up to 78 pairs of integral field unit (IFU) spectrographs in VIRUS spread out over the HET's focal plane with fill factor of 1:4.5.  Each IFU is fed by 448 fibers \citep{Hill+21} and each observation consists of 3 exposures of  6.1-minute duration that are dithered to fill in the coverage within the IFU footprints. The resulting spectra, $\sim 33$K per observation, are calibrated, sky subtracted, and scanned for emission lines (or continua) without any color or magnitude pre-selection, and, where lines (or continua) are found, the spectra from the surrounding fibers inside a 3\farcs5 radius aperture are combined into a single, point-spread function (PSF) weighted spectrum \citep{Gebhardt+2021}. \rvtwo{Though the apertures are large, for the  1\farcs7 average seeing FWHM, 90\% of the light for the spectrum comes from the inner-most 1\farcs2. }

The PSF weighted spectra are classified and assigned a redshift with the \textit{ELiXer} software \citep{Davis_2022}\footnote{The HPSC-1 classifications derive from an earlier version of the ELiXer software than is presented in \cite{Davis_2022}} using multiple analyses and incorporating archival photometric imaging. Brighter ($g<22$) spectra are classified with support from \textit{Diagnose}\footnote{\url{https://github.com/grzeimann/Diagnose}}, a software package developed for the Hobby-Eberly Telescope VIRUS Parallel Survey which is is based on template fitting \citep{Zeimann_2023}. Additional support is also provided from the \textit{Dark Energy Explorers}\footnote{\url{https://www.zooniverse.org/projects/erinmc/dark-energy-explorers}} citizen science project \citep[][]{House_2023}. 

For this work, we select the 51,863 HPSC-1 spectra from objects classified as $1.88 < z < 3.52$ galaxies, explicitly excluding objects containing Active Galactic Nuclei (AGN), as identified by ELiXer or \cite{Liu_2022a} \rv{(see also \S \ref{caveats})}.  The redshift range is defined by where \lya\ falls within the HETDEX spectral bandpass. \rv{The AGN classifications made by ELiXer within this redshift range are based on (1) the detection of \lya\ found with other emission lines combinations consistent with AGN, such as \ciii, \civ, \heii, etc, (2) broad \lya\ emission $\gtrsim$1200 \kms, or (3) a position and photometric magnitude match to an object from an external catalog that is classified as an AGN. More exhaustive AGN classifications are made by \cite{Liu_2022a} based on single emission line and line pair matching with enhanced, multi-Gaussian fitting, the continuum profile, and visual inspection of all candidate AGN HETDEX spectra along with archival photometry. }\\

\section{Methods} \label{sec:methods}

The details on the HETDEX data pipeline, including calibrations, \rv{sky subtraction}, and spectra extraction, are found in \cite{Gebhardt+2021}.  The creation and general description of the public data catalog release, HPSC-1, is described in \cite{Cooper_2022}. All galaxy spectra used in this work come from the aperture extracted, wavelength-rectified, PSF weighted, atmospheric dispersion corrected, Milky Way dust de-reddened data presented in HPSC-1. The additional corrections and operations applied to those spectra are described in the subsections that follow.\\

\rv{\subsection{Sky Subtraction} \label{sky_subtraction}}

\rv{Though the sky subtraction methodology is described in detail in \cite{Gebhardt+2021}, because of its direct relevance to to several topics in this work, we present a brief overview of the salient points. The goal of the sky subtraction is to remove background light while preserving the light from astronomical sources. The background light comes from the atmosphere (literally, the ``sky"), as well as instrumental effects such as thermal noise and scattered light in the optics. Furthermore, any large diffuse astrophysical source will be included in the sky background; for example, light from unresolved faint foreground and background galaxies that are uniformly distributed in the sky will add to the estimate of the background. There are additional complications discussed in \cite{Gebhardt+2021} such as wavelength distortion in the spectrographs that can lead to sky background residuals.}

\rv{Here we make use only of the \textit{local} form of the HETDEX sky subtraction, as it is more stable than the \textit{full field} sky subtraction in the pipeline version used to create the HPSC-1, the source of the spectra for this work. The on-sky size of the local sky is $50\arcsec\times12$\arcsec, whereas the full sky is derived from the entire 21\arcmin\ diameter focal plane. While both sky subtractions use similar methods, the local sky subtraction employs 112 fibers for each amplifier (4 per IFU), and the full field sky subtraction uses all fibers in an exposure ($\sim$ 35K). The full sky necessitates additional adjustments for amplifier to amplifier variations and other complications. For emission line source detection, we rely on the local sky since it is more robust to instrumental effects and generates smaller background residuals. }

\rv{For the local sky subtraction, fibers containing obvious continua are removed, including those with counts more than 3$\times$ the biweight scale of the fibers on the amplifier. The remaining fibers are then used to compute the per-wavelength bin background from their biweight locations. Typically, 70-80\% of the fibers in an amplifier are used for this calculation, though it can be as low as 25-30\% when there is a star or other bright object in or near the FoV of the amplifier. As HETDEX detections can include fibers from more than one amplifier, due to their size and/or their position within an IFU, the fibers included in their detection aperture can have different local sky subtraction corrections. One potential consequence of the local sky subtraction, which we revisit later, is that it can partly remove low-level flux actually associated with detected galaxies if that flux is spatially extended over a large portion of an IFU.}\\

\subsection{Sky Subtraction Residual} \label{residual_subtraction}

Similar to the Background Residual Correction presented in \cite{Davis_2021}, the LAE spectra stacked in this work have an observed-frame residual correction applied prior to their shift to the restframe. The purpose of this correction is to account for small systematics in the calibration, and the average extra light in the extraction apertures that come from faint, undetected background and foreground sources that are undetectable at the individual detection level. In \cite{Davis_2021}, apparently empty individual``Sky Fibers" are selected, sorted, and stacked on a per observation basis and used to correct the individual fibers of the LAE detection from the matched observation. However, in this work, given the large number statistics of the 250$\times$ larger LAE sample, we select entire apertures rather than individual fibers and average over many observations instead of applying separate corrections for each observation.  

Using the observations from June 2018, when the IFU array became significantly populated, through June 2020, the end of data acquisition for HPSC-1, we collect up to 200 spectra from random ``empty" $3\farcs 5$ radius apertures within each of the $\sim$ 2000 observations in this range; this radius matches the size of the standard HETDEX aperture \citep{Gebhardt+2021}, and the spectra are extracted using the same method as the normal HETDEX detections. The center of each aperture is randomly selected from the footprint of the observation's IFU coverage with the following constraints: 
\begin{enumerate}
    \item No aperture center can be within $1\farcs 5$ of another. This means the apertures can overlap but not near their central regions where the PSF weights are highest. 
    \item There must be at least 15 fibers included in the aperture over the three exposures comprising the observation. This avoids apertures where a significant portion falls off the edge of an IFU.
    \item The measured $g$ magnitude in the PSF weighted spectra extracted under the aperture must be fainter than 24, and have a median flux density between 3900~\angstrom~ and 5400~\angstrom~ greater than $-4 \times 10^{-18}$ \cgsa. This satisfies the ``empty" requirement, given the HETDEX depth is close to 25 in $g$ in good conditions but can be near 24 in poor seeing \citep{Gebhardt+2021,Cooper_2022,Davis_2022}. The negative lower limit accounts for the possibility of some over subtraction and for excursions due to noise.
    \item There can be no detected emission lines within the extracted spectrum and no HETDEX detections within $2\farcs 0$.
\end{enumerate}

\begin{figure}[ht]
    \centering
    \hspace*{-1.0cm}
    \includegraphics[width=0.5\textwidth]
    {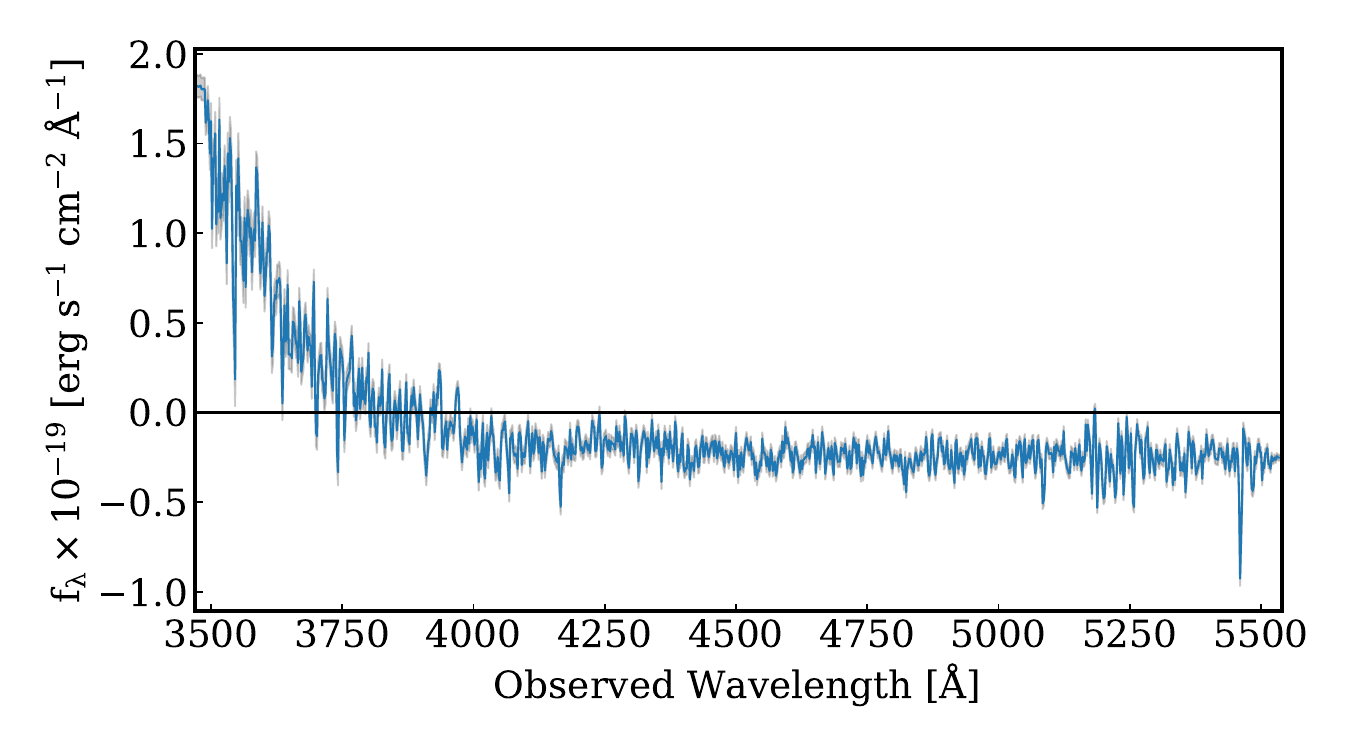}
    \caption{Weighted biweight averaged of the residuals produced by sky subtraction derived from stacking more than 300K random, ``empty" 3\farcs5 radius apertures taken from HPSC-1 HETDEX observations. The errors are shown in gray and are derived via $\sigma_{b}/\sqrt{N}$ where $\sigma_{b}$ is the biweight scale. The $f_{\lambda}$ level of this residual spectrum is $\sim$1\% of the HETDEX flux limits \citep{Gebhardt+2021} and thus meaningful only when stacking large numbers of spectra. The rise in the far blue end of the spectrum is primarily due to a limitation in the instrument and the slightly negative flux redward of $\sim$4000~\AA\ is a result of a small over-correction in the sky subtraction. The spectrum is available in electronic form.}
    \label{fig:background_residual}
\end{figure}

These criteria yield more than 300K PSF weighted spectra (not all observations yield the target 200 ``empty" apertures) that are stacked along their native HETDEX 2~\angstrom\ wavelength bins using the biweight measure of central location \citep{Beers_1990} in the same method described in \cite{Davis_2021}. The result is shown in Figure \ref{fig:background_residual} with the residual spectrum flux in blue and the flux uncertainties for each wavelength bin in gray. The reported uncertainties are statistical and are defined as $\sigma_{b}/\sqrt{N}$ where $\sigma_{b}$ is the biweight scale of the $N$ contributing ``empty" aperture measured spectra for that wavelength bin. The spectrum is also made available in electronic form. Sky subtraction residual spectra created by stacking under alternate sub-selections of the full $\sim$300K sample are stable with respect to variations in the seeing FHWM, throughput and instrument response, date, and observed declination. 

Each LAE spectrum is corrected by subtracting this sky subtraction residual in its observed frame. For individual spectra the correction is less than 1\%, far below the noise level; the systematic becomes meaningful only when stacking 100s or 1000s of galaxy spectra. Figure \ref{fig:background_residual} shows a distinct rise in the far blue caused by instrumental limitations and is slightly negative redward of about 4000~\angstrom\ indicating a small over-subtraction by the reduction pipeline. Both these effects are more than 100$\times$ smaller than the typical HETDEX flux limits \citep[$\sim2\times10^{-17}$ \cgsa, but can rise up to 10$\times$ higher blueward of 3800\AA; ][]{Gebhardt+2021} and are inconsequential for individual spectra. They only become significant when stacking large numbers of spectra. Future data releases will include improvements in the data reduction pipeline to better address these issues.\\

\subsection{Shift to Restframe}

 As described in the previous subsection, the spectra are first corrected for the sky subtraction residual in their observed frame. Depending on the needs of the analysis, all spectra are corrected for wavelength- and redshift-dependent fractional IGM transmission. The IGM correction applied in this work is from CIGALE \citep{CIGALE_2019}, which derives from \cite{Meiksin}. The individual application of the IGM transmission correction can be highly uncertain since IGM attenuation is dependent on the particular sight line \citep{Inoue_2014}. However, since we are averaging over many thousands of spectra, this becomes naturally statistical \citep{Dijkstra_2007,Laursen_2011,Zheng_2011,Behrens_2018,Steidel_2018,Gurung-Lopez_2019,Byrohl_2020}. Lastly, the spectra are shifted to their own restframe so that they can be aligned for stacking.
 
The restframe shift has two components, one for the wavelength bins and one for the flux density at each bin center. Since the reported HETDEX wavelengths are in air, the observed wavelengths are converted to vacuum \citep{Greisen_2006} and then simply adjusted by (1+$z$) where $z$ is determined from the Gaussian fitted line center of the \lya\ emission line \citep{Cooper_2022}. As discussed in \cite{Davis_2021} and later in this work (\S \ref{lya_velocity_offset}), there is no correction applied for any \lya\ velocity offset from an individual galaxy's systemic redshift. Depending on the science needs, this wavelength-only shift, by itself, may be sufficient, but it only represents the observed flux at rest wavelengths: when stacking, this can create a bias against higher redshift objects due to increased cosmological dimming. For this work, and to counteract this bias, we also convert the observed flux density ($f_{\lambda}$) to a luminosity density ($L_{\lambda}$ or $L_{\nu}$, as needed). Please note that here we use ``luminosity density" as an analog to flux density, not as a luminosity per unit volume. This conversion is a straightforward application of:
\begin{equation} \label{eq:flux_to_lum}
    L_{\lambda} = 4\pi~D_L^2~f_{\lambda}~(1+z),
\end{equation}
where $D_{L}$ is the luminosity distance. Where $L_{\nu}$ is needed, we simply multiply $L_{\lambda}$ by $\lambda^2$/c. Unless stated otherwise, $z$ (specifically, $1+z$) is defined as the ratio of the observed frame, vacuum-shifted Gaussian-fitted emission line center wavelength to the vacuum restframe wavelength. 
\\

\subsection{Stacking} \label{stacking}

Once the previously described corrections and shifts are made, the spectra are stacked using the same method described in \cite{Davis_2021}. The restframe wavelength spacing is adopted from the highest redshift object to be stacked, with the blue and red wavelength endpoints coming from the highest and lowest redshift objects respectively. All spectra are then linearly interpolated onto this wavelength grid and stacked in each wavelength bin using the weighted biweight statistic \citep{Davis_2021}, as a modification of the biweight statistic described in \cite{Beers_1990}. \rv{The weighted biweight alters the biweight location slightly by including an additional weight for each element in the sample. The weight used in the stack is the inverse of the uncertainty on the flux measures within each wavelength bin, such that the more uncertain fluxes have a reduced contribution to the biweight location.} We note that for these large datasets ($\gg$$10^4$ spectra), the biweight and weighted biweight perform very similarly to a median average, \rv{with a brief comparison of the two methods provided in \S \ref{sec:analysis}. In all cases, the default tuning constants} \rvtwo{are used. For the biweight location (or weighted biweight location) the constant is 6 and for the biweight scale it is 9.}

While it is common practice to normalize the spectra before stacking, often to the flux near 1500~\AA\ for LBGs and LAEs, \citep[][and others]{Vargas_2014, Marchi_2017, Steidel_2018}, the nature of the HETDEX data makes this impractical. For the vast majority of the spectra presented in this work, only the \lya\ emission is detectable above the flux limits, so there is no measured continuum against which to normalize. Instead, we stack the un-normalized spectra first and then, where appropriate for the analysis, we normalize the stack against the flux over a section of its wavelengths.

Since the HETDEX wavelength window is fixed at 3500-5500~\AA\ and the LAEs span $1.9 < z < 3.5$ \citep{Gebhardt+2021}, only the wavelength region immediately around rest-frame \lya\ can receive contributions from all spectra (the number of contributing spectra for each wavelength bin is included as a column in the stacked spectrum data file). The fluxes in wavelength bins increasingly blueward of \lya\ are populated by LAEs at increasingly higher redshifts, while the opposite is true moving redward of \lya. \\

\section {Results} \label{sec:analysis}

\begin{figure*}[ht]
    \centering
    \includegraphics[width=1\textwidth]{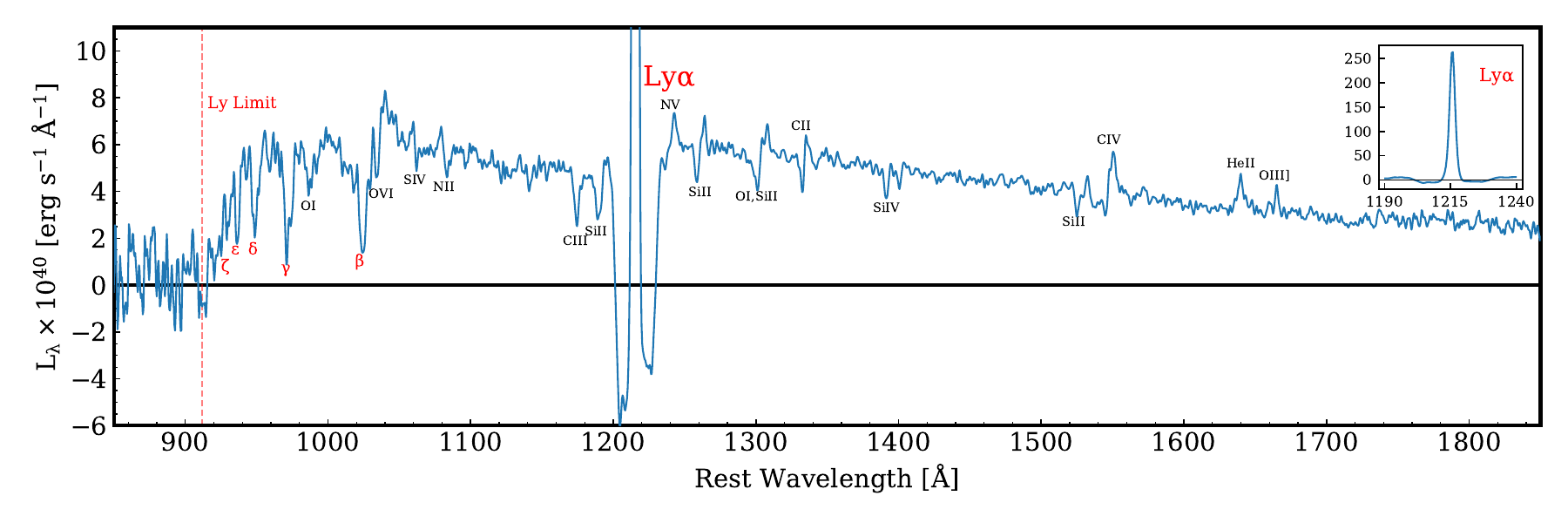}
    \caption{Restframe weighted biweight-averaged stack of $\sim$50K LAEs from HPSC-1 between $1.9 < z < 3.5$ (not corrected for IGM transmission). The spectrum has been smoothed with a 1 pixel (0.44~\AA) Gaussian kernel ($\sigma$). As noted in \S \ref{stacking}, the number of contributing galaxies \rv{(Figure \ref{fig:smooth_stack_2panel} green curve, bottom panel)} decreases with distance from \lya, dropping to 16\% in the extreme blue and 4\% in the red. The uncertainties \rv{(Figure \ref{fig:smooth_stack_2panel} orange curve, top panel)} are $\sim$5\% except near edges of the wavelength range where they rise to $\sim$25\% at the red end and $\sim$100\% in the far blue. Despite containing more contributing spectra, the blue end of the spectrum has larger uncertainties due to the performance of the CCD \citep{Gebhardt+2021}. These uncertainties are defined as $\sigma_{b}/\sqrt{n_{\lambda}}$ where $\sigma_{b}$ is the biweight scale and $n_{\lambda}$ is the number of spectra contributing to the wavelength bin. The inset in the upper right shows the full height of the \lya line in the same luminosity units. \rv{The deep troughs near the \lya\ line have some physical motivation, but are artificially enhanced by the data reduction pipeline (see \S \ref{lya_troughs}).} This spectrum, with the associated error array, is available in electronic form.} 
    \label{fig:smooth_stack}
\end{figure*}

The stack of the full 50K $1.9 < z < 3.5$ LAEs (median $z$ = 2.55), not corrected for IGM transmission, is presented in Figure \ref{fig:smooth_stack} in terms of $L_{\lambda}$ normalized to the median between 1475 and 1525~\AA\ ($L_{1500}$). The data for this figure, including the number of contributing spectra and uncertainties for each wavelength bin, is available in electronic form.  

\begin{figure}[ht]
    \centering
        \hspace*{-1.0cm}
    \includegraphics[width=0.5\textwidth]{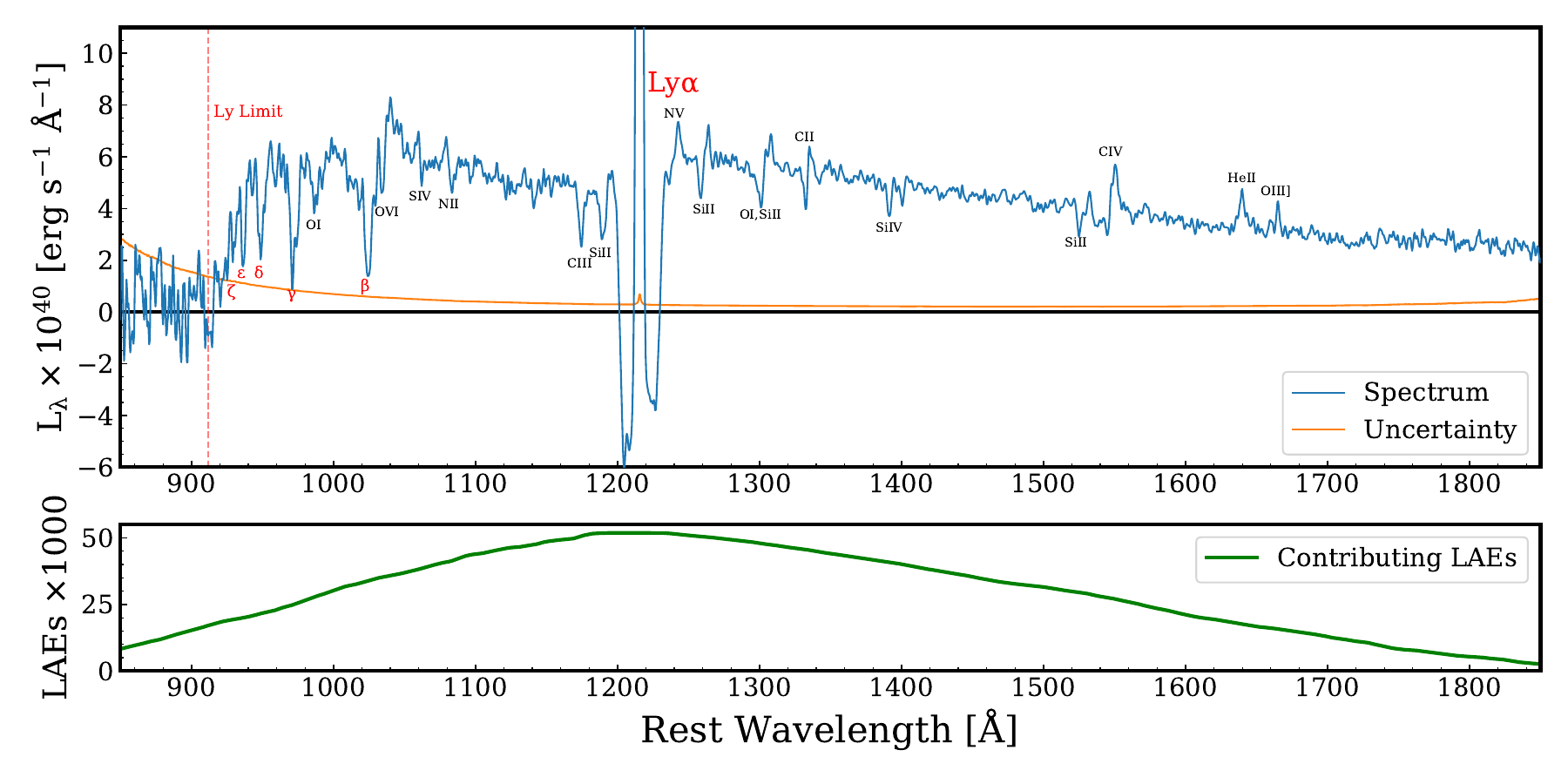}
    \caption{\rv{Restframe weighted biweight-averaged stack of $\sim50$K LAEs from HPSC-1 between $1.9 < z < 3.5$ (not corrected for IGM transmission), as in Figure \ref{fig:smooth_stack}, with the uncertainty shown in in orange (top panel) and the number of contributing galaxies per wavelength bin in green (bottom panel).} }
    \label{fig:smooth_stack_2panel}
\end{figure}

\begin{figure}[ht]
    \centering
        \hspace*{-1.0cm}
    \includegraphics[width=0.5\textwidth]{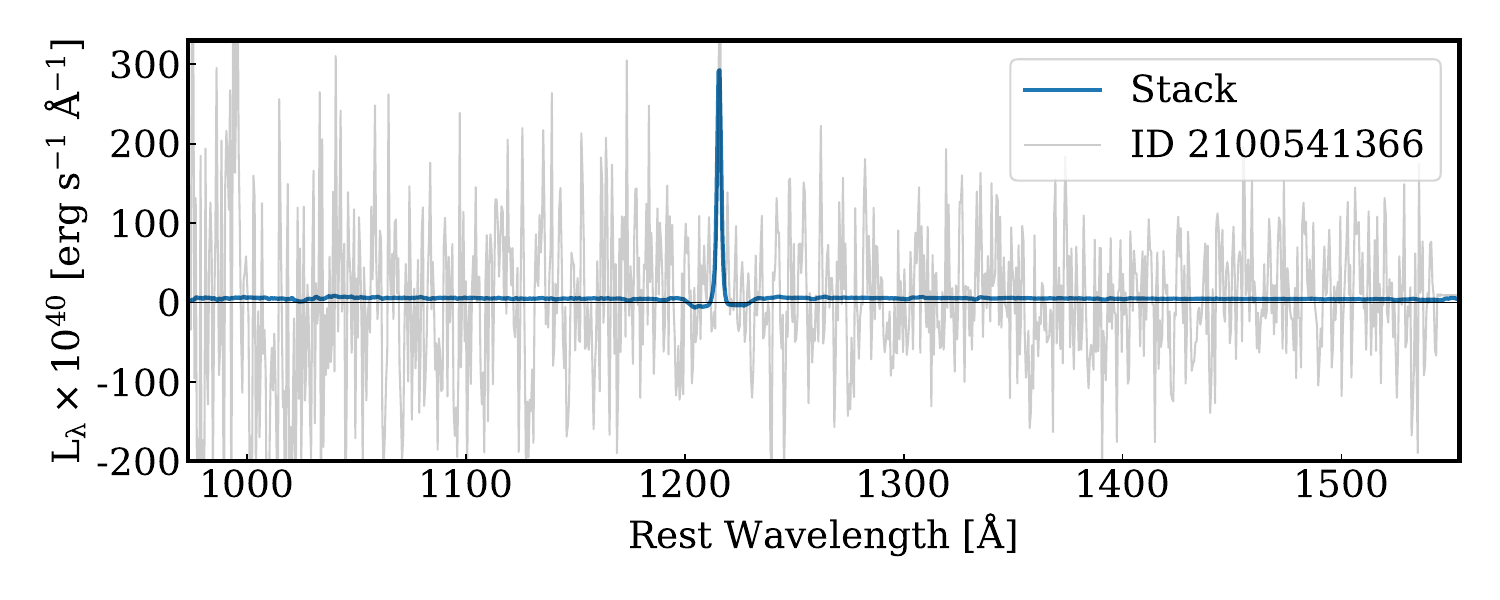}
    \caption{Stack from Figure \ref{fig:smooth_stack} compared to a typical HPSC-1 LAE, ID: 2100541366 (RA,Dec: 150.127594 +2.295267) to illustrate the signal-to-noise gains. This LAE is selected as it has properties similar to the sample average and ensemble: $z$ = 2.566, $g$ = 25.3, \lya\ SNR = 5.9, \lya\ log Luminosity = 42.81 (ergs~s$^{-1}$), $W_{\lambda}$(\lya) = 87~\AA\null. The figure is cropped to the restframe wavelengths covered by the individual detection and aligned at \lya. If we define the SNR simply as the ratio of the means of the flux and the errors on that flux \citep[][and \S \ref{sec:analysis} in this work]{Gebhardt+2021} between 1250 and 1650~\AA, we see an improvement of more than 300$\times$ from the \rv{continuum} SNR $\approx$ 0.06 in the individual detection to the SNR $\approx$ 21 in the stack. }
    \label{fig:stack_v_det}
\end{figure}

The \lya\ emission line fitting is performed with ELiXer \citep{Davis_2022} and the SNR has increased from a median of 6 for the individual galaxies in the sample to near 1000 in the stack. The uncertainties on the stack (\rv{shown as the orange curve in the top panel of Figure \ref{fig:smooth_stack_2panel} as absolute uncertainties }) are $\sim$5\%, except near the wavelengths farthest from \lya, where they grow to $\gtrsim$25\% in the far red and spike above 100\% in the far blue.  The latter effect is due to the declining CCD sensitivity in the blue \citep{Gebhardt+2021} and the decreasing number of contributing objects \rv{(shown as the green curve in the bottom panel of Figure \ref{fig:smooth_stack_2panel})} at the spectral extremes (16\% of the maximum in the far blue and 4\% in the far red). These uncertainties are defined like standard errors with $\sigma_{b}/\sqrt{n_{\lambda}}$, where $\sigma_{b}$ is the biweight scale and $n_{\lambda}$ is the number of spectra contributing to the wavelength bin centered on $\lambda$. Figure \ref{fig:stack_v_det} compares the stack to a typical HPSC-1 LAE (ID: 2100541366; RA,Dec: 150.127594 +2.295267), selected for the similarity of its properties ($z$ = 2.566, $g$ = 25.3, \lya\ SNR = 5.9, \lya\ log Luminosity = 42.81 (ergs~s$^{-1}$), and $W_{\lambda}$(\lya) = 87~\AA) to the ensemble and sample averages. The figure illustrates the SNR improvements gained through stacking. Using wavelengths redward of \lya\ (1250-1650~\AA, rest) and defining the SNR simply as the mean flux divided by the mean (analogous) error on that flux \citep{Gebhardt+2021} and the standard error-like definition above, there is an increase of more than $300\times$ from the mean SNR $\approx$ 0.06 for the continuum in an individual detection to the mean SNR $\approx$ 21 in the stack.

 \rv{As noted in earlier, the differences between the biweight and median methods are small for large sample sizes. For the spectra in this work, the \textit{median} of the differences between stacks using the biweight method (\S \ref{stacking} and Figure \ref{fig:smooth_stack}) and using a median is 2.6\%, measured from restframe 850\AA\ to 1850\AA, or 3.1\% over the entire wavelength range, 768\AA\ to 1918\AA. For the same stacks, the \textit{biweight location} of the differences are nearly identical to the \textit{median} differences at 2.6\% and 3.0\% for the same two wavelength ranges. Here we define the difference per wavelength bin as:
\begin{equation} \label{abs_diff_bw_md}
     \frac{\mid L_{\lambda}^b - L_{\lambda}^m\mid}{\frac{1}{2}(L_{\lambda}^b + L_{\lambda}^m)},
\end{equation}
where $L_{\lambda}^b$ is the luminosity in wavelength bins from the biweight "averaged" stack and $L_{\lambda}^m$ is the luminosity in the same wavelength bins from the median``averaged" stack (see the METHODS section (\S \ref{sec:methods}) for the details on the stacking processes). For Gaussian distributions, the biweight location and median differ by $\ll$ 1\% and the standard deviation, $\sigma$, and the biweight scale, $\sigma_{b}$, by $<$ 1\%.}

Selected features of the stack (Figure \ref{fig:smooth_stack}) are discussed and interpreted, quantitatively and qualitatively, in subsequent subsections.\\

\subsection{Caveats} \label{caveats}

Before proceeding further, it is necessary to acknowledge several caveats that can affect the analyses and interpretations. 

Contamination of the sample of HPSC-1 LAEs by misclassified emission, in particular that of \oii, is quite low at an estimated value of $\sim$2.5-3.0\% \citep{Cooper_2022}.  However, since it is not zero, it may slightly dilute the signal of the underlying continuum.  Another 0.5-2.0\% of contamination could come from other individually detected lines such as \ciii, \civ, or \MgII, some of which are associated with possible AGN. And though we make efforts to exclude AGN through the identification of broadline ($>$ 1200 km s$^{-1}$) pairs (e.g., \lya\ + \civ, \civ + \ciii, \ovi + \lya, etc), emission line profile fitting, catalog matching, and visual inspection \rv{\citep{Liu_2022a,Davis_2022}}, some AGN (particularly single narrow-line Type-II AGN and some broadline objects where a second line is not identified by the pipeline) may remain in the LAE sample \citep{Cooper_2022,Davis_2022,Liu_2022a,LujanNiemeyer_2022}. Some fraction of the detections, particularly at lower emission line SNR, may also be false detections of noise. While work is on-going to better quantify the specific rates of these false detections, for the LAEs of HPSC-1, as described in \S 6.6 of \cite{Cooper_2022} positively confirms 91\% of the detections in a sub-sample of LAEs with repeat observations and thus sets an upper limit of 9\% on the total fraction of contaminants. 

Unlike \cite{Davis_2021}, the spectra in the stack have not been selected to exclude LAEs with nearby neighbors, as that information is absent from this data release.  Similarly, the spectra have not been deblended to remove the contribution of flux by neighboring objects \citep{Davis_2023b}. While the removal of the sky subtraction residual should largely handle the average contribution of faint line-of-sight interlopers, excess flux from these sky-adjacent neighbors can still be present in the individual galaxy spectra and find its way into the stacks.

While the HETDEX catalog has no galaxy preselection, its emission line detections are flux limited \citep[$\sim4\times10^{-17}$ \cgs; ][]{Gebhardt+2021}, and the HPSC-1 catalog excludes detections with emission line SNR $< 5.5$ \citep{Cooper_2022}. There is thus a bias against detecting galaxies with observationally fainter \lya, which necessarily increases with increasing redshift. However, the use of the median-like weighted biweight mechanics \citep{Beers_1990,Davis_2021} in the stack does mitigate the influence of bright outliers, which helps maintain the representative nature of the stack.

As previously stated and further discussed in \S \ref{lya_velocity_offset}, the spectra are aligned for stacking based on their Gaussian fitted \lya\ line. This can lead to a $\sim$1\AA\ smearing of spectral features due to the individual \lya\ redshift offsets from the galaxies' systemic redshifts. 

Lastly, as noted in \S \ref{stacking},  the wavelength regions farther from \lya have fewer contributing spectra and thus increased uncertainty in the stack compared to wavelengths closer to the \lya\ line. Furthermore, we have to assume that evolution in the galaxies over those redshift ranges of each stack is minor, in order to consider the stacked spectrum, as a whole, as representative of the underlying sample. \\

\subsection{Representative Stack} \label{sec:representative_stack}

\begin{figure}[ht]
    \centering
    \hspace*{-1.0cm}
    \includegraphics[width=0.5\textwidth]{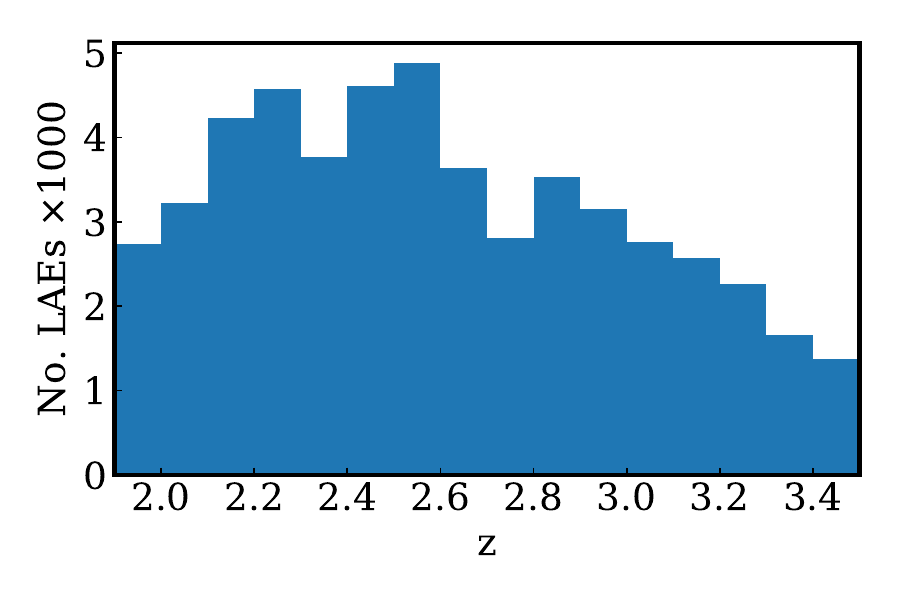}
    \caption{The redshift distribution (mean $z=2.6$) of the $\sim$50K, flux-limited detection LAEs with \lya\ SNR $>$ 5.5 from the HPSC-1.}
    \label{fig:laes_by_z_hist}
\end{figure}

Table \ref{tab:summary_table} presents a summary of several properties (see the table note for descriptions) of the full dataset and full stack along with three additional subselections that divide the sample by redshift. \rvtwo{The redshift distribution, with mean $z=2.6$, for this work is shown in Figure \ref{fig:laes_by_z_hist}.} Additional binnings based on other properties will be presented in future works. With the exception of the average $r$ magnitude, which is included from various archival photometric imaging catalogs that overlap the HETDEX observations \citep{Davis_2021,Davis_2022}, the properties are measured from the individual spectra or the stacked spectra. A comparison of the \rv{biweight location ($\tilde{x_b}$)} values of $g$, \lya\ Luminosity, and \lya\ equivalent width to the values derived from the stacked spectra demonstrates that they agree very well, suggesting the stacks are, indeed, similar to an \rv{average} representation of the samples. We acknowledge that both the \rv{biweights} of $g$ and $W_{\lambda}$(\lya) are more limited comparisons as HETDEX cannot measure continuum magnitudes fainter than $\sim$25 in $g$, hence $W_{\lambda}$(\lya) is a lower limit \citep{Gebhardt+2021,Cooper_2022,Davis_2022}.\\

\begin{deluxetable*}{c|c|c|c|c|c|c|c|c}[ht]
\tablecaption{\rv{Summary of Ensemble Properties by Redshift Range} \label{tab:summary_table}}
\tablewidth{0pt}
\tablehead{ 
\colhead{Redshift}  &
\colhead{N} &
\colhead{$\langle g\rangle$} &
\colhead{Stack $g$}  &
\colhead{$\langle r\rangle$} &
\colhead{$\langle L_{Ly\alpha}\rangle$} &
\colhead{Stack $L_{Ly\alpha}$} &
\colhead{$\langle W_{\lambda}$(\lya)$\rangle$} &
\colhead{Stack $W_{\lambda}$(\lya)}  \\
\colhead{(1)} &
\colhead{(2)} &
\colhead{(3)} &
\colhead{(4)} &
\colhead{(5)} &
\colhead{(6)} &
\colhead{(7)} &
\colhead{(8)} &
\colhead{(9)}  
}
\startdata 
$1.9 < z < 3.5$  & 52K & $\tilde{x_b}$=25.9, $\sigma_b$=1.30 &  25.6 & $\tilde{x_b}$=25.5, $\sigma_b$=1.16& $\tilde{x_b}$=42.92, $\sigma_b$=0.212 & 42.83 &  $\tilde{x_b}$=93.1, $\sigma_b$=84.0 & 114.0   \tabularnewline
\hline
$2.0 < z < 2.5$ & 21K & $\tilde{x_b}$=25.8, $\sigma_b$=1.35  &  25.3 & $\tilde{x_b}$=25.3, $\sigma_b$=1.21 & $\tilde{x_b}$=42.86, $\sigma_b$=0.211 & 42.77 & $\tilde{x_b}$=89.9, $\sigma_b$=86.5 & 108.8  \tabularnewline
$2.5 < z < 3.0$ & 18K & $\tilde{x_b}$=25.8, $\sigma_b$=1.27  &  25.6 & $\tilde{x_b}$=25.4, $\sigma_b$=1.12 & $\tilde{x_b}$=42.91, $\sigma_b$=0.187 & 42.84 & $\tilde{x_b}$=80.9, $\sigma_b$=72.6 & 101.7 \tabularnewline
$3.0 < z < 3.5$ & 11K & $\tilde{x_b}$=26.2, $\sigma_b$=1.18  &  26.2 & $\tilde{x_b}$=26.0, $\sigma_b$=0.83 & $\tilde{x_b}$=43.03, $\sigma_b$=0.180 & 42.96 &$\tilde{x_b}$=107, $\sigma_b$=81.4 & 130.0  \tabularnewline
\enddata
\tablecomments{\small{
$^{(1)}$ Redshift range of (sub)selection.
$^{(2)}$ Number of galaxies in the (sub)selection to the nearest 1000.
$^{(3)}$The biweight location ($\tilde{x_b}$) and biweight scale ($\sigma_b$) of the SDSS-$g$ magnitude of individual LAE detections as computed from the HETDEX spectra using the \textit{speclite} Python package (\citealp{speclite_pkg}; \citealp[see also][]{Davis_2022}). Though always computed, values  $\gtrsim$25 are fainter than the HETDEX detection limit.
$^{(4)}$The SDSS-$g$ magnitude computed from the observed frame stack of HETDEX spectra, again using \textit{speclite}. Given the high SNR of the stack, the formal error on the fit magnitude is $<$ 0.01.
$^{(5)}$The biweight location and biweight scale of the $r$ magnitudes of individual LAE detections as computed from photometric imaging with $r$ coverage. Depth is catalog dependent \citep{Davis_2022}, with 75\% of HPSC-1 $\gtrsim$26. Non-detections in the imaging are included as their respective limits.
$^{(6)}$The log$_{10}$ of the biweight location and biweight scale of the luminosity [erg s$^{-1}$] of the \lya~ line of the individual LAE detections.
$^{(7)}$The log$_{10}$ of the luminosity [in erg s$^{-1}$] of the Gaussian fitted \lya line of the stack. Given the high SNR, uncertainties on the fits are $\ll$1\% 
$^{(8)}$The biweight location and biweight scale of the restframe equivalent width [\angstrom] of \lya in the individual LAE detections. Since the continuum is often undetected, this is a lower limit.
$^{(9)}$The restframe equivalent width [\angstrom] of \lya\ in the stack. Given the high SNR, the errors on the fit are $ < 0.05~\angstrom$.
}}
\end{deluxetable*}

\subsection{Lyman Alpha Velocity Offset} \label{lya_velocity_offset}

Since we align the individual spectra based on their fitted, restframe \lya emission line centers, our figures show the stacked \lya line centered at our adopted wavelength of 1215.67~\AA\null. However, due to the complexities of \lya\ radiative transfer and the suppression of the flux near and just blueward of restframe \lya\ \citep[][and many others]{Verhamme_2006,Verhamme_2018,smith_2018,Byrohl_2020}, we are often fitting to the red peak of \lya, and there is an offset with respect to each galaxy's systemic redshift. In the stack, this manifests as a slight offset between the expected and observed positions of the other emission and absorption features. Additionally, as the \lya\ offset from systemic is variable by galaxy, there can also be a smearing/broadening of these other (stacked) spectral lines. With only \lya\ detected in the vast majority of the HETDEX LAE spectra, and generally at SNR $\lesssim$6 for the HPSC-1, our ability to correct for the velocity offset of individual galaxies prior to stacking is limited \citep{Davis_2021}. However, as we estimate the typical velocity offset to be at most a few hundred km s$^{-1}$ (see below), the impact to this work is small and we do not refine it further here. 

To estimate the average \lya\ velocity offset in our ensemble, we repeatedly fit the center lines of several emission and absorption features using a simple Markov Chain Monte Carlo (MCMC) Gaussian fitting code and compute a velocity offset from the assumed fiducial wavelength. These features are selected as they are clear, have high SNR and are not significantly blended with any other lines (or can have the blended portion easily masked). The results are summarized in Table \ref{tab:velocity_offsets}.

 The velocity offsets from the features are similar, though there is some obvious scatter likely as a combination of ISM and IGM confusion, outflows, and other kinematics. The overall mean, 235 $\pm$ 18 km s$^{-1}$, provides a good estimate of the \lya\ velocity offset for stack and thus for the typical HPSC-1 LAE. Where stated for the remainder of this work, we adopt a rounded value of 250 km s$^{-1}$ for the velocity offset from systemic for the stack. \rv{Though LAEs can exhibit a sizeable difference in individual \lya\ velocity offsets, this average is consistent with those of LAEs and LBGs found in \citet{Shapley_2003} ($\sim$ 360 \kms), \citet{Erb_2014} ($\sim$ 240 \kms), \cite{Shibuya_2014} ($\sim$ 230 \kms), \citet{Steidel_2018} ($\sim$ 300 \kms), and \citet{Muzahid_2020} ($\sim$ 170 \kms).} A more rigorous investigation is presented in \cite{Weiss_2023}.

 
These estimates generally agree with and bracket the $\sim$200 km s$^{-1}$ reported for the much smaller, $z > 3$ sample in \cite{Davis_2021}.  This equates to a less than 1\AA\ offset in the adopted Lyman Continuum region, 880-910\AA. That work finds no significant impact to the Lyman Continuum estimate measured over 30\AA\ and no apparent change when applying a correction such as that in \cite{Verhamme_2018,Byrohl_2019,Gurung_L_pez_2020}. This small velocity offset is also consistent with the expectation of enhanced Lyman Continuum leakage \citep{Izotov_2018,Naidu_2021}.\\

\begin{deluxetable}{l|c|c}[ht]
\tablecaption{Lyman Alpha Velocity Offsets} \label{tab:velocity_offsets}
\tablewidth{0pt}
\tablehead{
\colhead{Line}  & \colhead{Fiducial $\lambda$ [\AA]} & \colhead{Offset [km s$^{-1}$]} 
}
\startdata 
Ly$\beta$\tablenotemark{\footnotesize ~(a)} & 1025.72 & 320 $\pm$ 58  \tabularnewline
\ion{C}{3} & 1175.71  & 330 $\pm$ 37  \tabularnewline
\civ \tablenotemark{\footnotesize ~(b)} & 1549.48 & 199 $\pm$ 27 \tabularnewline
\heii\tablenotemark{\footnotesize ~(c)} & 1640.42 & 125 $\pm$ 41  \tabularnewline
\ion{O}{3}]\tablenotemark{\footnotesize ~(d)} & 1666.15 & 198 $\pm$ 28  \tabularnewline
\hline
mean & NA & 235 $\pm$ 18
\enddata
\tablecomments{Velocity offsets of various spectral lines from the \lya\ aligned HPSC-1 stack (Figure \ref{fig:smooth_stack}) as measured against the MCMC Gaussian fitted line centers. The last row is a simple, unweighted mean. Scatter is likely due to a combination of ISM and IGM confusion, outflows, and other kinematics.}
\tablenotetext{a}{ subject to ISM and IGM confusion}
\tablenotetext{b}{ doublet; $\lambda$ as unweighted mean}
\tablenotetext{c}{ shows stellar (broad) and nebular (narrow) components}
\tablenotetext{d}{ doublet; using the red peak only}
\end{deluxetable}

\subsection{Lyman Alpha Troughs} \label{lya_troughs}

\rvtwo{On either side of the \lya\ emission line there are deep, negative ``troughs." While absorption is expected near the \lya\ emission, the depths of the troughs, as shown in Figure \ref{fig:smooth_stack}, are enhanced by our reductions. Though scattering of photons near \lya\ resonance by the \ion{H}{1} in and around the LAEs is expected, the majority of the depths of these troughs are a result of the HETDEX sky subtraction and reduction pipeline (\S \ref{sky_subtraction}). The sky subtraction assumes that the sky off-source is the same as the sky on-source. Any differences in this assumption could then create a feature in the resulting stacked spectra. As one possible scenario, the version of the pipeline in this data release could over-subtract faint \lya\ emission from halos around the LAEs that extend $\sim$10 or more arcseconds. In this case, we could create a self subtraction of the \lya\ emission and cause the troughs to go negative. Another possible scenario is that a diffuse UV background exists as part of the general sky background we measure. In this case, the on-source sky could be different than the off-source sky due to absorption around \lya. A more complete discussion of the artificial troughs and real \lya\ scattering is presented in \cite{Weiss_2023}. For the measurements of this work, we simply avoid the \lya\ troughs by masking them.}

\subsection{Lyman Alpha Luminosity} \label{sec_lya_lum}

The calculations of the \lya\ luminosity for both individual galaxy spectra and the stack are similar. For an individual galaxy, the identified \lya\ emission line is fit with a simple Gaussian \citep{Davis_2022} whose area is the integrated flux, noting that the continuum level is a free parameter and allowed to be negative. That flux is then converted to a luminosity using Eq.~\ref{eq:flux_to_lum}, but without the tailing $(1+z)$ term. For the $L_{\lambda}$ stacked spectra, since we are already in the rest frame and in luminosity density, we simply fit a Gaussian where the continuum is set by the $L_{\lambda}$ redward of \lya\ and whose area is the integrated line luminosity with the \lya\ trough regions (\S \ref{lya_troughs}) masked out. This is the observed luminosity of the escaping \lya. There are no additional corrections applied for extinction or attenuation by the restframe dust and ISM.  

\begin{figure}[ht]
    \centering
    \hspace*{-1.0cm}
    \includegraphics[width=0.5\textwidth]{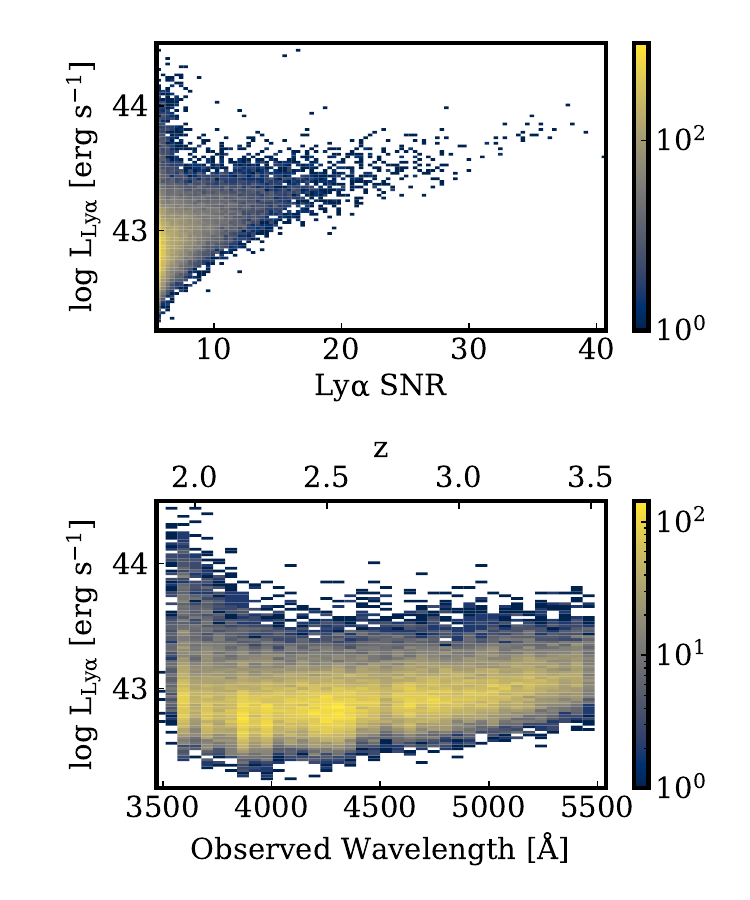}
    \caption{\lya\ luminosity in the HPSC-1 as functions of \lya\ SNR and observed wavelength. \rv{The color, independently for each panel, tracks the number of galaxies.} The highest luminosities occur at the lowest SNR and bluest wavelengths where the diminishing throughput of the instrument amplifies the scatter intrinsic to low SNR detections. The lower figure also shows a trend to larger luminosities with longer wavelengths; this is due to the HETDEX flux limits \citep{Gebhardt+2021} and \rv{is consistent with} cosmological dimming.}
    \label{fig:lum_vs_wave_vs_snr}
\end{figure}

As shown in Table \ref{tab:summary_table} and Figure \ref{fig:lum_vs_wave_vs_snr}, the individual galaxy \lya\ log luminosities (i.e., log$_{10}$($L_{Ly\alpha}$/[erg s$^{-1}$])) in this sample range from 42.26 to 44.45 with a \rv{biweight location} of 42.92 $\pm$ 0.212 and 1.9\% of the \lya\ emission lines greater than 43.50. At the brighter end, it is probable that some of the galaxies host unidentified AGN and some luminosities may be inflated due to measurement uncertainty. The full-sample stacked spectra has a \lya\ luminosity of 42.83, about 20\% lower than the sample \rv{biweight, though well within the uncertainty}. \rv{The fit to the \lya\ line of the stack} is a more \rv{precise ``average"} of the sample as a whole \rv{due to the increased SNR, since the continuum level is a free parameter for both the stack and the individual detections.} The stacked continuum is well detected \rv{where continuum is rarely detected in individual HETDEX LAEs and the greatest spread in luminosities in the sample is associated with the lowest \lya\ signal-to-noise detections, which represent the largest fraction of the sample (Figure \ref{fig:lum_vs_wave_vs_snr})}.

Table \ref{tab:summary_table} naively shows a weak trend of increasing \lya\ luminosity with increasing redshift bin, which could indicate some small evolution with redshift \citep[][for example]{Ciardullo_2012}. \rv{However, the decrease is consistent with the loss due to the HETDEX flux limits and cosmological dimming.}\\

\subsection{Lyman Alpha Equivalent Widths} \label{sec_lya_ew}

\begin{figure}[ht]
    \centering
    \hspace*{-1.0cm}
    \includegraphics[width=0.5\textwidth]{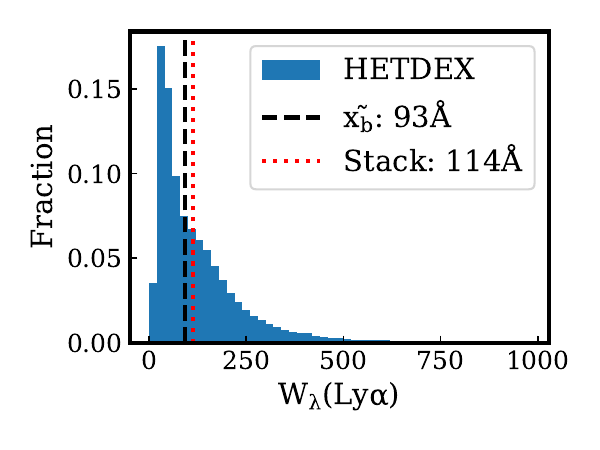}
    \caption{Distribution of \lya\ restframe equivalent widths in HPSC-1. The y-axis is normalized to the full 50K LAEs of this work. The black dashed vertical line marks the \rv{biweight location} value of the distribution and the red dotted line the value as computed from the stack of the spectra. As most LAEs in HPSC-1 do not have detected continua, the individual equivalent widths are generally lower limits. The continuum in the stacked spectrum (Figure \ref{fig:smooth_stack}) is clearly detected and that computed equivalent width is more secure. See also Table \ref{tab:summary_table}. }
    \label{fig:ew_hist}
\end{figure}

The \lya\ restframe equivalent widths, $W_{\lambda}$(\lya), are not explicitly reported in the HPSC-1 but are shown in Figure \ref{fig:ew_hist}, along with the sample \rv{biweight location} and the corresponding stack value. Each $W_{\lambda}$(\lya) is computed from the integrated \lya\ flux and a combined estimation of the continuum from the spectrum and associated photometry \citep{Cooper_2022,Davis_2022} divided by (1+$z$). For each of the restframe stacked spectra (Table \ref{tab:summary_table}), $W_{\lambda}$(\lya) is simply the ratio of the integrated \lya\ luminosity to the fitted continuum around the \lya\ line, again with the \lya\ troughs masked. As with the \lya\ luminosities, the stacked spectra measurement of $W_{\lambda}$(\lya) is consistent with the \rv{biweight} measure from the sample, but the stack \rv{may} provide a more robust average as the continuum is detected; HETDEX rarely detects the continua of $z \gtrsim 1.9$ galaxies.

\begin{figure}[ht]
    \centering
    \hspace*{-1.0cm}
    \includegraphics[width=0.5\textwidth]{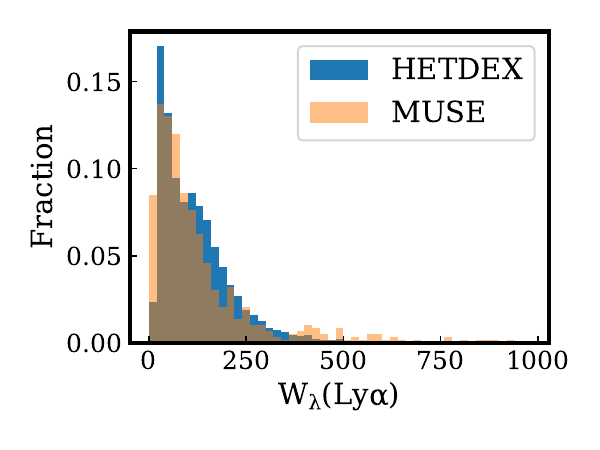}
    \caption{Normalized histograms of the 13.5K HPSC-1 LAEs in $2.9 < z < 3.5$ and the 591 MUSE LAEs of the same redshift range \citep{Kerutt_2022}. The distributions are similar, \rv{with 11\% of these HETDEX LAEs and 16\% of the MUSE LAEs exhibiting estimated $W_{\lambda}$(\lya) values $>240\angstrom$ and with biweight locations of $W_{\lambda}$(\lya) values of 102.7 $\pm$ 80.2~\AA\ and 85.4 $\pm$ 78.7~\AA\ respectively.}}
    \label{fig:ew_hist_vs_MUSE}
\end{figure}

The HPSC-1 equivalent width distribution is similar to that found with MUSE in \cite{Kerutt_2022}. HPSC-1 reports that 15\% of its LAEs have $W_{\lambda}$(\lya) $>240~\angstrom$ and the full sample has a \rv{biweight location value of 93.1 $\pm$ 84.0~\angstrom}; \cite{Kerutt_2022} reports that 16\% (in their full sample) of the galaxies have $W_{\lambda}$(\lya) $>240\angstrom$ and the characteristic $W_{\lambda}$(\lya) is 95.5~\angstrom.  \rv{Restricting the comparison of the two samples to their LAEs within the overlapping redshift coverage, $2.9 < z < 3.5$, the distributions deviate a bit more, but remain similar, see Figure \ref{fig:ew_hist_vs_MUSE}. The redshift restricted MUSE sample of 591 LAEs (out of 1920 LAEs of the full sample) has 16\% with  
$W_{\lambda}$(\lya) above 240~\AA\ vs 11\% for the redshift restricted HPSC-1 sample of 13.5K LAEs. The biweight locations of the restricted MUSE $W_{\lambda}$(\lya) and HPSC-1 samples are 85.4 \AA\ and 102.7 \AA, respectively.}

\begin{figure}[ht]
    \centering
        \hspace*{-1.0cm}
    \includegraphics[width=0.5\textwidth]{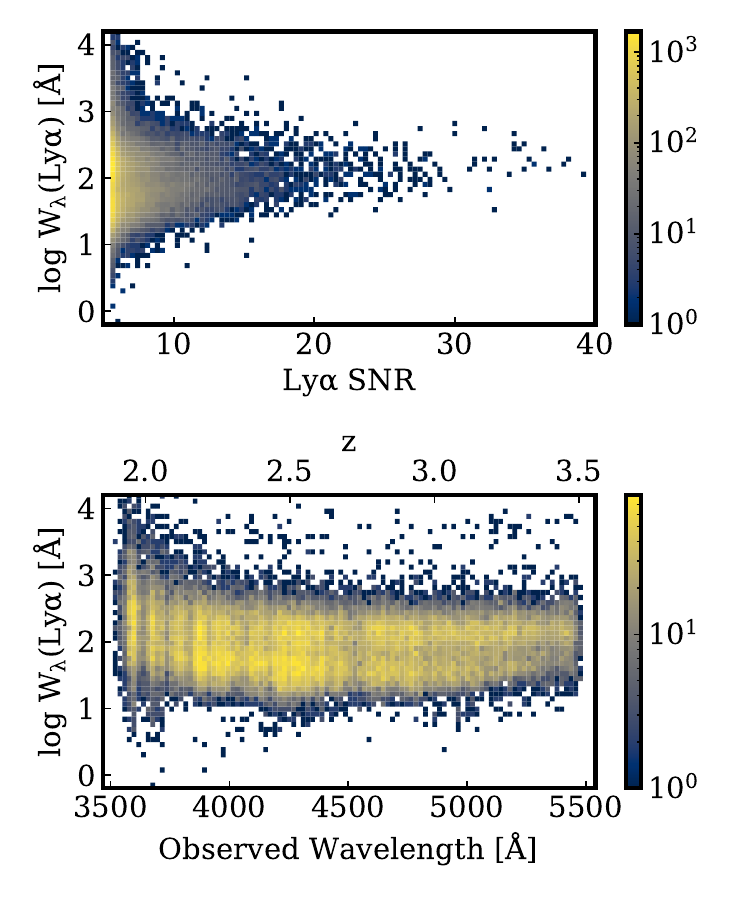}
    \caption{\lya\ restframe equivalent widths in the HPSC-1 as functions of \lya\ SNR and observed wavelength. \rv{The color, independently for each panel, tracks the number of galaxies.} Since most HPSC-1 spectra do not have a detectable continuum, the equivalent widths are lower limits. There is an obvious loss of lower $W_{\lambda}$(\lya) objects with increasing redshift, which is likely just the result of the HETDEX flux limits \rv{and cosmological dimming}.  The evidence for a true increase in equivalent width with redshift in this dataset is \rv{very weak at best}. See also Table \ref{tab:summary_table}.}
    \label{fig:ew_vs_wave_vs_snr}
\end{figure}

The HPSC-1 equivalent widths given in Table \ref{tab:summary_table} and Figure \ref{fig:ew_vs_wave_vs_snr} might also suggest some weak evolution with redshift similar to findings in \cite{Ciardullo_2012}.  However, this is even less clear with the HPSC-1 data than the possible trend with \lya\ luminosity as the equivalent width measures are less secure due to the lack of detected continua and are also biased toward the detection of brighter \lya\ at higher redshifts \rv{due to cosmological dimming}.\\

\subsection{UV Continuum Slope and UV Magnitude} \label{uvb_slope}

The observed UV continuum slope, $\beta$, modeled as $f(\lambda) \propto \lambda^{\beta}$, is taken from the full stack (Figure \ref{fig:smooth_stack}) and measured from 1250~\AA\ to 1850~\AA\ by fitting a simple least squares optimized power law after masking the features near 1260, 1302, 1335, 1394, 1549, 1640, and 1665~\AA\null. Given the wavelength range, this is similar to $\beta_{18}$ in \cite{Calzetti_2001}. The best fit $\beta$, $-2.36 \pm 0.09$, implies that our ``average" LAE has young, low metallicity stellar population, a large ionizing photon production efficiency, and very little dust extinction \citep[][]{Calzetti_2001,Bouwens_2010b,Calabr_2021,Chisholm_2022,saldana-lopez_2022}. This is in line with that expected for the $z \sim 3$ LAE population (see also \S \ref{ficus}). This also suggests an increased likelihood of Lyman Continuum escape and may indicate a recent (5-15 Myr) burst of star formation \citep[][]{Calzetti_2001,Calabr_2021,Chisholm_2022,saldana-lopez_2022}.

The absolute UV Magnitude (M$_{\text{UV}}$ or M$_{\text{1500}}$) of the full stack is computed from the mean and median luminosity density between 1400 and 1600~\AA\ in the restframe, \rv{using the standard 3631 Jy AB magnitude zero point scaled to $L_{\nu,0}$ of 4.3454$\times$10$^{20}$ erg s$^{-1}$ Hz$^{-1}$ at 10 pc. The M$_{\text{UV}}$ from the mean values corresponds to a medium bright galaxy with $-19.67 \pm$ 0.13, while the median values are essentially identical, $-19.68 \pm$ 0.13. The errors are statistical and come from the propagation of the standard deviation of the flux density between 1400 and 1600~\AA\ in the restframe.} \\

\subsection{P Cygni Profiles} \label{sec_pcygni}

P Cygni line profiles \citep{Beals_1934} are clearly visible in Figure \ref{fig:smooth_stack}, for example \ion{O}{6} (1032,1038~\AA), \NV (1241~\AA), and, to a lesser degree, \CIV (1549~\AA). These profiles are complex combinations of nebular and stellar absorption and emission, and contain a wealth of information on the stellar population, star formation history, and the metal enrichment of the galaxy, but a proper decomposition and fitting is beyond the intended scope of this work. However, qualitatively speaking, the P Cygni profile is a tell-tale indicator of strong stellar winds, massive stars, and recent star formation \citep[][]{Steidel_1996,Leitherer_2001,Chisholm_2019}. While not exhibiting a P Cygni, the extremely broad \heii (1640~\AA) emission, 940 $\pm$ 130 km s$^{-1}$, is also strongly suggestive of a very young population \citep{Chisholm_2019}.   This is consistent with our picture of LAEs, and the clarity of the profile in the stack further highlights the SNR gains and additional physics that is not directly accessible in the individual HPSC-1 spectra.
\\

 \subsection{SED Fitting} \label{ficus}

\begin{figure*}[ht]
    \centering
    \includegraphics[width=1\textwidth]{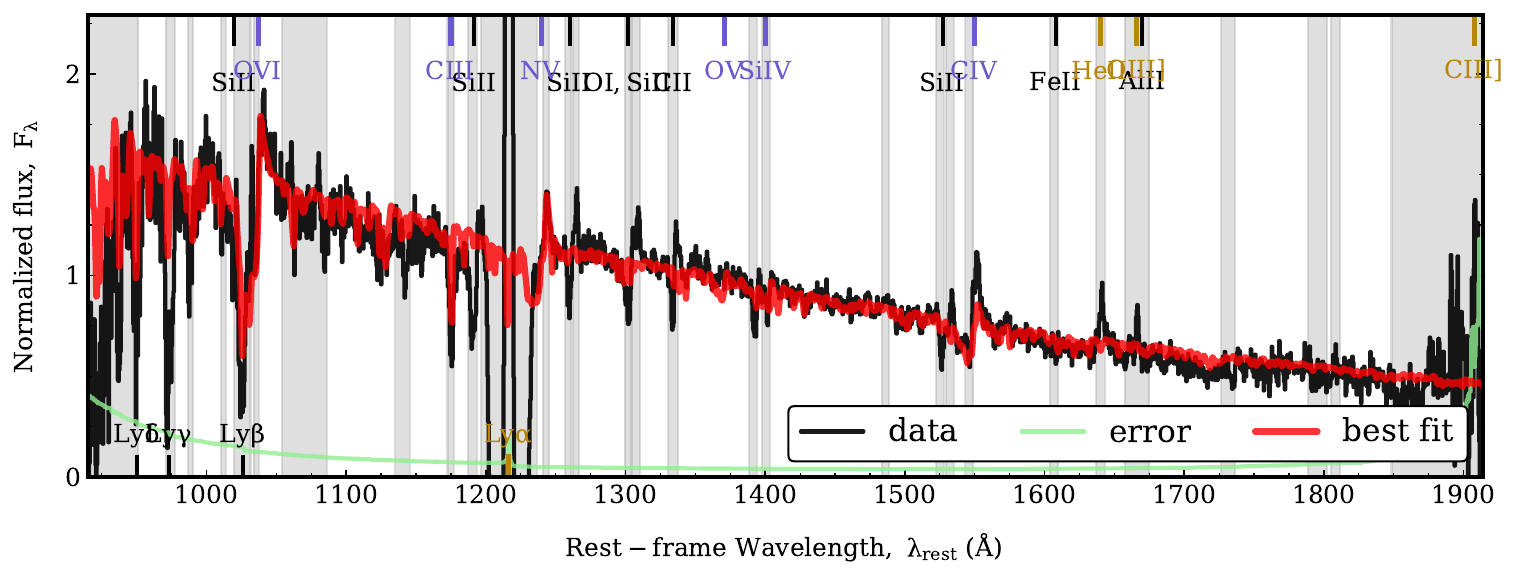}
    \caption{Best fitting, lowest-$\chi^{2}$, FiCUS model (SB99+SMC in Table \ref{tab:ficus_summary_table}). The black curve is the IGM corrected version of the full 50K stacked spectrum shown in Figure \ref{fig:smooth_stack} and the red curve is the FiCUS stellar continuum fit. Light gray regions, which correspond to most of the spectral features, are masked and not fit by the code.  FiCUS is not designed to fit interstellar absorption or emission features.} 
    \label{fig:ficus_top}
\end{figure*}

To model and gain some understanding of the underlying stellar population, we perform Spectral Energy Distribution (SED) fitting on the stacked spectrum (Figure \ref{fig:smooth_stack}). Since we are limited by the wavelength coverage in this spectrum to the restframe UV, we are only probing more recent star formation. The SED fitting is performed with the Python package for \textit{\textbf{Fi}tting the stellar \textbf{C}ontinuum of \textbf{U}V \textbf{S}pectra} or FiCUS\footnote{\url{https://github.com/asalda/FiCUS}} \citep{saldana-lopez_2022}. Four separate runs use the Starburst99 \citep[SB99;][]{Leitherer_2010} and the Binary Population and Spectral Synthesis \citep[BPASS, v2.2.1;][]{Eldridge_2017} single-burst stellar population models, along with the dust attenuation models from \citet[][hereafter, R16]{Reddy_2016} and the SMC extinction model by \citet{Prevot_1984}. All runs assume the same \citet{Kroupa_2001} Initial Mass Function (IMF) with a 100 M$_\odot$ upper limit and fit for ten stellar ages (1, 2, 3, 4, 5, 8, 10, 15, 20, and 40 Myr) and four metallicites (5\%, 20\%, 40\%, and 100\% of Z$_{\odot}$). We use the IGM corrected version of the full stack spectrum (Figure \ref{fig:smooth_stack}) with an applied velocity offset of 250 km s$^{-1}$ (\S \ref{lya_velocity_offset}) and fit over 915 - 1915~\AA\ in the rest-frame, assuming a mean redshift of 2.604 for the 50K contributing galaxies. Key results of the 4 runs are summarized in Table \ref{tab:ficus_summary_table} with the fit with the best $\chi^2$ (row 2, SB99 + SMC) shown in Figure \ref{fig:ficus_top}. We are not modelling Lyman Continuum escape with FiCUS, as we are only fitting the continuum redward of the Lyman Limit and are not yet confident using the restricted wavelength range for the $z > 3.0$ HETDEX LAEs where Lyman Continuum could be observed.

While there are some small differences in the results from the four runs, with BPASS favoring slightly older, less enriched stellar populations and the R16 dust model favoring slightly increased reddening, a consistent picture emerges of an ensemble-averaged galaxy with a young (10-15 Myr), metal poor (0.2-0.3 Z$_{\odot}$) stellar population with minimal (E(B-V) 0.03 - 0.10) reddening. The observed and intrinsic UV continuum slopes (ranging $\beta = -2.10$ to $-2.05$ and $-2.65$ to $-2.54$, respectively) for the four runs are fit over 1250 - 1850~\AA\ as described in \S \ref{uvb_slope} and bracket the $-2.36$ UV continuum slope measured directly from the HPSC-1 stack. \cite{saldana-lopez_2022}, with support from \cite{Izotov_2016}, \cite{Salim_2018}, and \cite{Shivaei_2020}, suggests that SB99 + SMC may best represent the conditions for these LAEs and indeed, though perhaps coincidentally, that combination does produce the lowest $\chi^2$ fit. This model (SB99 + SMC) does favor the youngest, UV light-weighted average stellar age with a significant fraction, just over 75\%, of the light coming from stars with ages less than 5 Myr. \\

\begin{deluxetable*}{c|c|c|c|c|c|c|c}[ht]
\tablecaption{Summary of FiCUS SED Fitting} \label{tab:ficus_summary_table}
\tablewidth{0pt}
\tablehead{
\colhead{Model}  & \colhead{$\chi^2$} & \colhead{Age (Myr)} & \colhead{$<$ 5 Myr} &  \colhead{Z (Z$_{\odot})$} &  \colhead{E(B-V)} & \colhead{UV$\beta$} & \colhead{UV$\beta^\mathrm{Int}$}   
}
\startdata 
SB99 + R16 & 1.80 &  12.06 $\pm$ 1.41 & 0.75 & 0.26 $\pm$ 0.02 & 0.099 $\pm$ 0.002 & -2.05 $\pm$ 0.01 & -2.61 $\pm$ 0.01 \tabularnewline
SB99 + SMC & 1.77 & 10.86 $\pm$ 1.41 & 0.75 & 0.28 $\pm$ 0.02 &  0.037 $\pm$ 0.001 & -2.05 $\pm$ 0.01 & -2.65 $\pm$ 0.01 \tabularnewline
BPASS + R16  & 1.80   &  15.68 $\pm$ 0.88 & 0.35 & 0.21 $\pm$ 0.02 & 0.078 $\pm$ 0.003 & -2.10 $\pm$ 0.01 & -2.54 $\pm$ 0.01  \tabularnewline
BPASS + SMC & 1.78   &  14.79 $\pm$ 1.00 & 0.40 & 0.22 $\pm$ 0.02 & 0.030 $\pm$ 0.001 & -2.07 $\pm$ 0.01 & -2.57 $\pm$ 0.01   \tabularnewline
\enddata
\tablecomments{Results of four independent runs of the FiCUS stellar continuum-SED fitting code (\S \ref{ficus}). Errors are statistical only. \textbf{$\chi^2$} is the goodness of fit to the HPSC-1 stacked spectrum (Figure \ref{fig:smooth_stack}). \textbf{Age(Myr)} is the model UV light-weighted average stellar age. \textbf{$<$5 Myr} is the fraction of the fitted stellar population younger than 5 Myr. \textbf{Z} is the model UV light-weigthed average metallicity as a fraction of Solar metallicity (\textbf{Z$_{\odot}$}). \textbf{UV$\beta$} is the UV continuum slope of the observed flux FiCUS model between 1250 and 1850~\AA\ described in \S \ref{uvb_slope}. \textbf{UV$\beta^{int}$} uses the same fitting but applied to the FiCUS intrinsic flux model.}
\end{deluxetable*}

\subsection{Relative Lyman Continuum Escape and Comparison to Reference Sample} \label{lyc_escape}

For this measurement, we use $L_{\nu}$ instead of $f_{\nu}$ as explained in \S \ref{stacking}. We sub-select the 11K HPSC-1 LAEs with $z > 3$, where the restframe Lyman Continuum region, defined as [880 - 910~\AA] for consistency with literature \citep[][and many others]{Shapley_2006,Marchi_2017,Steidel_2018}, falls within the HETDEX spectral range. The $L_{900}$ value used is the weighted biweight location of the \rv{68 wavelength bin luminosities} in the residual subtracted (\S \ref{residual_subtraction}) and IGM transmission corrected \citep{CIGALE_2019} $L_{\nu}$ spectrum between 880 and 910~\angstrom. \rv{The reported error is statistical; a standard error analog as the biweight scale of the same data divided by $\sqrt{\textrm{n}}$, where n = 68. }

Since 1500~\angstrom\ is not available in the HETDEX spectral window for $z > 3$ galaxies, the $L_{1500}$ value is estimated using two methods. The first estimate is based on a scaling of the weighted biweight flux density between 1268 and 1296~\angstrom\ using the slope fit between 1250 and 1525\angstrom\ from our $2.6 < z < 3.5$ stack of 24K LAEs. That wider redshift range is selected as the narrowest window that includes sufficient wavelength coverage to the red of \lya\ to capture the $L_{1500}$ region. The $L_{1500}$ normalized fitted slope is 5.11 ($\pm$ 0.561)$\times10^{-4}$ L$_{\nu}$/\AA, and results in a stack average IGM corrected $L_{900}/L_{1500}$(\textit{out}) = 0.169 $\pm$ 0.0361 (or $L_{900}/L_{1500}$(\textit{obs}) = 0.069 $\pm$ 0.0137, without the IGM correction) as an upper limit, given the aforementioned caveats. \rv{The labels, \textit{out} and \textit{obs}, adopt the convention in \citet{Steidel_2018,pahl_2021,pahl_2022}}. The second $L_{1500}$ estimate simply uses the weighted biweight \rv{luminosity} density \rv{of the 113 wavelength bins between 1475 and 1525~\AA)} taken directly from the $2.6 < z < 3.5$ stack \rv{in the same method as $L_{900}$ described earlier.} This result differs by $\ll 1\%$, completely consistent with the first estimate. \rv{Repeating the same computation, but substituting the median and standard deviation for the weighted biweight location and biweight scale, yields an IGM transmission corrected $L_{900}/L_{1500}$(\textit{out}) = 0.159 $\pm$ 0.0455, or $L_{900}/L_{1500}$(\textit{obs}) = 0.066 $\pm$ 0.0172 when not correcting for the IGM.}

\begin{figure*}[ht]
    \centering
    \includegraphics[width=1\textwidth]{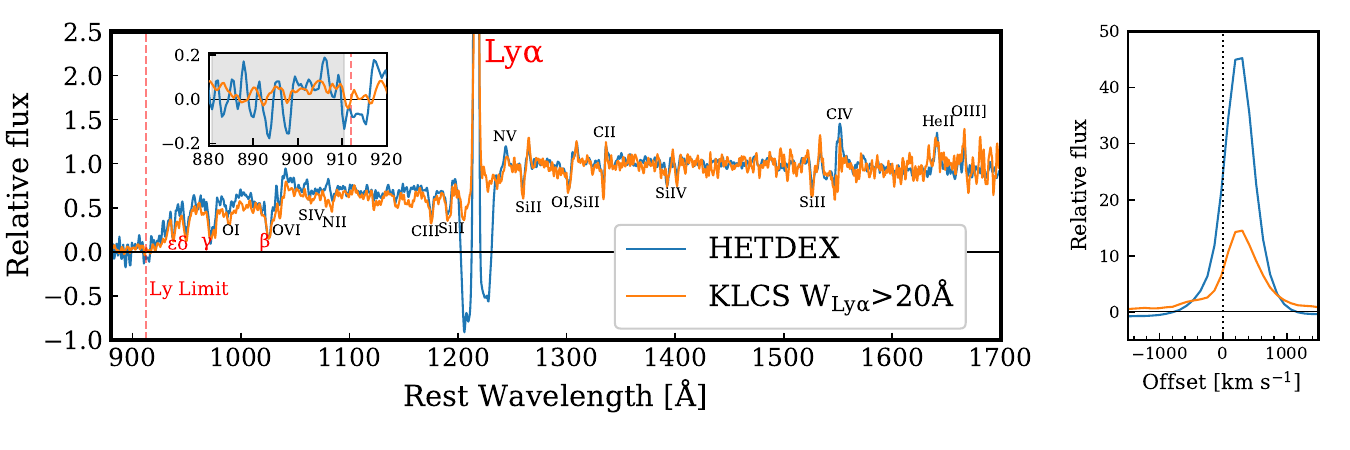}
    \caption{Comparison of the HETDEX (HPSC-1) restframe stack to stack of the classical LAE subset (26 out of 124) of KLCS with $W_{\lambda}$(\lya) $>$ 20\AA~ \citep{Steidel_2018,pahl_2021}.
    The HPSC-1 stack has been corrected for IGM transmission, shifted by 250 km s$^{-1}$ (\S \ref{lya_velocity_offset}), and converted to $L_{\nu}$ units to match the KLCS data. Both spectra have been interpolated to the same wavelengths and smoothed with a 1 pixel (0.44~\AA) Gaussian kernel ($\sigma$). The left-hand panel shows a section of the UV restframe. \rv{The zoom-in insert shows the Lyman Continuum region (shaded).} The right-hand panel shows a zoom around the \lya\ lines in velocity space relative to the assumed systemic redshift. 
   }
    \label{fig:vs_klcs_20AA}
\end{figure*}

To place this measurement in some context, we compare against the stack of the subset (26 out of 124 galaxies) of the LBG selected, $z \sim 3$ galaxies in the Keck Lyman Continuum Spectroscopic Survey (KLCS) \citep{Steidel_2018} that are also classical LAEs with ($W_{\lambda}$(\lya) $>$ 20\AA~ \citep{pahl_2021}. The overplotted stacks are presented in Figure \ref{fig:vs_klcs_20AA}. Both stacks are normalized to their own flux near 1500\AA~ and interpolated onto the same wavelength grid and smoothed with a 1-pixel (0.44~\AA) Gaussian kernel ($\sigma$). The HETDEX (HPSC-1) stack has also been shifted to correct for its approximate 250 km s$^{-1}$ \lya\ velocity offset (\S \ref{lya_velocity_offset}). Though the individual spectra contributing to the KLCS stack are of higher SNR with longer exposures, they have similar resolving powers, with $R\sim800$ for HETDEX and for KLCS at $\lambda < 5000\angstrom$ and $R\sim1400$ for KLCS at $\lambda > 5000\angstrom$ \citep{Steidel_2018,Gebhardt+2021}. Our $L_{900}/L_{1500}$ estimates are defined in a similar way as the $<$$f_{900}/f_{1500}$$>_\mathrm{out}$ measurements of \cite{Steidel_2018,pahl_2021} and others. 

\rv{As a brief aside, the \lya\ troughs (\S \ref{lya_troughs}) shown in the HPSC-1 stack are much deeper than in the KLCS stack. The KLCS LAE subsample is comprised of UV bright, LBG selected galaxies with weaker \lya, as compared to the majority of the 50K HPSC-1 LAEs. This difference in selection may probe galaxies with somewhat different halo and internal properties which could contribute to some differences in the troughs. However, as discussed in \cite{Weiss_2023} and briefly in \S \ref{lya_troughs}, while the existence of the \lya\ troughs is physically motivated, their manifestation within the HPSC-1 stack is substantially enhanced by the HETDEX data reduction pipeline.}

While our relative Lyman Continuum values, $L_{900}/L_{1500}$(\textit{out}), are $\sim1.5-2.0\times$ higher than what is found in \cite{pahl_2021,pahl_2022} for the galaxies in their highest ($W_{\lambda}$(\lya) $>$ 20\angstrom) equivalent width bin, based on \rv{a comparison to} the properties of the HPSC-1 sample, we might expect a larger escape of Lyman Continuum photons. \rv{Additionally, the $z \ge 3$ for $\sim$900~\AA\ restframe IGM transmission in the model used in this work \citep{CIGALE_2019} is about 10\% higher than that in \citet{Steidel_2018, pahl_2021}, making the IGM correction here slightly smaller.} As can be seen in Figure \ref{fig:vs_klcs_20AA}, the equivalent width of the HPSC-1 \lya\ line is roughly 3$\times$ larger than that of the KLCS LAE subsample, with HPSC-1 measuring 114\AA~ (or 130\AA~ for the $3.0 < z < 3.5$ subsample, see Table \ref{tab:summary_table}) and the KLCS LAE subsample measuring 36\AA\, using the same MCMC fitting code in \S \ref{sec_lya_ew}.  Measuring the UV continuum slope as in \S \ref{uvb_slope}, we find the KLCS subsample stack with a shallower, but still very blue, $\beta$ = -1.88 vs the -2.36 of the HPSC-1 stack. The FiCUS analyses for the KLCS subsample stack, with the same configurations as in \S \ref{ficus}, suggest a \rv{similarly young stellar population component with similar metallicity but a somewhat} larger E(B-V) as compared to the HPSC-1 sample. Using the SB99+SMC configuration, though all 4 runs show the same relative differences, we see the UV light-weighted average stellar age = \rv{9.36 $\pm$1.39} Myr, Z(Z$_\odot$) = \rv{0.21 $\pm$ 0.02}, and E(B-V) = \rv{0.052 $\pm$ 0.001}.

\rv{Both the KLCS and the HPSC-1 LAEs show consistent ages for their recent star formation as well as comparable metallicities, suggesting similar ionizing photon production. However, the stronger \lya\ emission, lower E(B-V) and steeper UV$\beta$ slope in the HPSC-1 sample promotes the expectation of increased leakage of those photons from the HPSC-1 LAEs} \citep{Behrens_2014,Verhamme_2015,smith_2018,Naidu_2021,Calabr_2021,Chisholm_2022,Maji_2022,saldana-lopez_2022}.

We emphasize that, while this relative $L_{900}/L_{1500}$ measurement can be used in context with other works, it is only a step in obtaining an estimate of the average, intrinsic escape fraction of ionizing photons from these galaxies. We also caution again (see \S \ref{caveats}) that the HPSC-1 sample is not as strictly controlled for contamination from on-sky neighbors as in \cite{pahl_2021} and that may influence the results. \rv{The subtraction of the average ``empty" aperture (\S \ref{residual_subtraction}) helps compensate for contributions of light from faint interlopers, but there is no exclusion of LAEs with detected, sky-adjacent neighbors as in \citet{Davis_2021} nor is there any de-blending of light from such neighbors as is performed in \citet{Davis_2023b}. As such, this result is an upper limit.} A more complete and extended study of the Lyman Continuum escape from $z > 3$ LAEs will be presented in \cite{Davis_2023b}.\\

\section{Summary}

We have taken the 50K low spectral resolution (R$\sim$800), generally low SNR LAE spectra (\lya\ SNR $\sim$6, continuum SNR $\ll$1) from the HETDEX Public Source Catalog 1 \citep{Cooper_2022}, applied a $\sim$1\% level correction for residual light, and stacked those spectra in the restframe using the weighted biweight method.  This procedure increased the spectral SNR by factors of several hundred and revealed a variety of otherwise noise obscured features associated with the LAEs. In stacking these large numbers of spectra, we marginalize over lines of sight, IGM transmission, galaxy orientation, and star formation stochasticity to yield a generally robust description of the ``average" or typical member of the set, though at the loss of peculiar features of individual members. The \lya\ luminosity (\S \ref{sec_lya_lum}), equivalent width (\S \ref{sec_lya_ew}), and $g$ magnitude of the stack (and redshift based substacks), are consistent with the corresponding median values of the LAE distribution, supporting the view of the stack as a valid ``average" representation (\S \ref{sec:representative_stack}).

The stack shows negative, asymmetric troughs (\S \ref{lya_troughs}) to either side of the \lya\ emission line. While real physics is behind the existence of the troughs, they are artificially enhanced by the HETDEX data reduction pipeline and are excluded in the analyses of this work. \cite{Weiss_2023} will explore the physics of the troughs and the HETDEX pipeline updates that address them.

The HETDEX LAE stack is a bit bluer with stronger \lya\ emission and less dust than the continuum selected LAE stack from the KLCS (\S \ref{lyc_escape} and Figure \ref{fig:vs_klcs_20AA}), but overall, is remarkably similar. 

We find that the properties of stacked spectra show our ``average" $z \sim 2.6$ LAE is very blue (UV$\beta$ $\sim -2.4$) with a significant light contribution from a young, metal poor stellar population (most of the UV light from stars with ages in 5-15 Myr, Z $\sim$ 0.2 Z$_{\odot}$, with strong P Cygni profiles and weak metal absorption). The steep UV$\beta$, low dust attenuation (E(B-V) $<$ 0.1), strong \lya\ emission (log $L_{Ly\alpha}$ $\sim$ 42.8, $W_{\lambda}$(\lya) $\sim$ 114\AA), and substantial $L_{900}$/$L_{1500}$(\textit{out}) ($\lesssim$ 17\%) all suggest a high intrinsic escape fraction of ionizing radiation.  This supports the idea that the higher redshift analogs of the HETDEX LAEs could be major drivers of Reionization.

Forthcoming research will expand and improve on these analyses with larger and more carefully curated LAE samples that will allow for finer binning and better exploration of the ensemble properties and their evolutions.

\acknowledgments

HETDEX is led by the University of Texas at Austin McDonald Observatory and Department of Astronomy with participation from the Ludwig-Maximilians-Universit\"at M\"unchen, Max-Planck-Institut f\"ur Extraterrestrische Physik (MPE), Leibniz-Institut f\"ur Astrophysik Potsdam (AIP), Texas A\&M University, The Pennsylvania State University, Institut f\"ur Astrophysik G\"ottingen, The University of Oxford, Max-Planck-Institut f\"ur Astrophysik (MPA), The University of Tokyo, and Missouri University of Science and Technology. In addition to Institutional support, HETDEX is funded by the National Science Foundation (grant AST-0926815), the State of Texas, the US Air Force (AFRL FA9451-04-2-0355), and generous support from private individuals and foundations.

Observations were obtained with the Hobby-Eberly Telescope (HET), which is a joint project of the University of Texas at Austin, the Pennsylvania State University, Ludwig-Maximilians-Universit\"at M\"unchen, and Georg-August-Universit\"at G\"ottingen. The HET is named in honor of its principal benefactors, William P. Hobby and Robert E. Eberly.

VIRUS is a joint project of the University of Texas at Austin, Leibniz-Institut f\"ur Astrophysik Potsdam (AIP), Texas A\&M University (TAMU), Max-Planck-Institut f\"ur Extraterrestrische Physik (MPE), Ludwig-Maximilians-Universit\"at Muenchen, Pennsylvania State University, Institut fur Astrophysik G\"ottingen, University of Oxford, and the Max-Planck-Institut f\"ur Astrophysik (MPA). In addition to Institutional support, VIRUS was partially funded by the National Science Foundation, the State of Texas, and generous support from private individuals and foundations.

The authors acknowledge the Texas Advanced Computing Center (TACC) at The University of Texas at Austin for providing high performance computing, visualization, and storage resources that have contributed to the research results reported within this paper. URL:http://www.tacc.utexas.edu

The Institute for Gravitation and the Cosmos is supported by the Eberly College of Science and the Office of the Senior Vice President for Research at the Pennsylvania State University.

KG acknowledges support from NSF-2008793.  EG acknowledges support from NSF grant AST-2206222. ASL acknowledges support from Swiss National Science Foundation. 
SS acknowledge the support for this work from NSF-2219212. SS is supported in part
by World Premier International Research Center Initiative (WPI Initiative), MEXT, Japan.

This research benefits from the open-source projects Python \citep{pythonref}, astropy \citep{astropy}, numpy \citep{harris2020array}, photutils \citep[][]{photutils}, and others in the open-source community.\\


\clearpage

\bibliography{hpsc1_lae.bib}
\clearpage

\end{document}